\documentclass[12pt,epsfig]{article} 
\input{epsf.tex}
\usepackage{amssymb,epsfig}
\newcommand{\One}{1\kern-4.5pt1}
\newcommand{\lapprox}{\raisebox{-0.5ex}{$\ 
\stackrel{\textstyle<}{\textstyle\sim}\ $}}
\begin{document}

\addtolength{\baselineskip}{0.20\baselineskip}

\rightline{SWAT/03/370}
\rightline{DUKE-TH-03-236}

\hfill February 2003


\vspace{48pt}

\centerline{\bf FERMI SURFACE PHENOMENA IN THE (2+1)$d$}
\centerline{\bf FOUR-FERMI MODEL} 


\vspace{18pt}

\centerline{\bf Simon Hands$^a$, John B. Kogut$^b$, 
Costas G. Strouthos$^{a,c}$ and Thao N. Tran$^b$}

\vspace{15pt}

\centerline{$^a$ {\sl Department of Physics, University of Wales Swansea,}}
\centerline{\sl Singleton Park, Swansea SA2 8PP, U.K.}
\smallskip

\centerline{$^b$ {\sl Department of Physics, University of Illinois at
Urbana-Champaign,}} 
\centerline{\sl 1110 West Green Street, Urbana, IL61801, U.S.A.}
\smallskip

\centerline{$^c$ {\sl Department of Physics, Duke University,}}
\centerline{\sl Durham, North Carolina 27708, U.S.A.}

\vspace{24pt}


\centerline{{\bf Abstract}}

\noindent
{\narrower 
We study the Gross-Neveu model in 2+1 dimensions with a baryon chemical
potential $\mu$ using both analytical and numerical methods.  For $\mu$ greater
than a critical value the model is chirally symmetric and has a Fermi surface
with $k_F\simeq\mu$. We have calculated the particle interaction in medium 
due to
scalar meson exchange to leading order in $N_f^{-1}$, where $N_f$ is the
number of flavors, in the hard dense loop approach. The result has been
used to calculate the relation between $\mu$ and the Fermi momentum and velocity
in the resulting Fermi liquid to $O(N_f^{-1})$. Simulation results from a
$32^2\times48$ lattice for fermion and meson dispersion relations and 
meson wavefunctions are then presented, showing qualitative and in some
cases quantitative 
agreement with analytic predictions. In particular, the simulations show clear
evidence for the in-medium modification of the scalar propagator, oscillatory
behaviour in the wavefunction consistent with a sharp Fermi surface, and
tentative evidence for a massless pole in the vector meson channel
resembling zero sound.
}




\vfill
\newpage
\section{Introduction}

Interest in QCD at low temperature
with a non-zero background baryon density is currently at a high level.
Understanding the ground state of any strongly interacting many-body system
requires non-perturbative techniques, and lattice QCD is
usually considered one of the most systematic and successful.
Unfortunately, attempts to develop Monte Carlo simulation techniques for 
systems with a non-vanishing baryon chemical potential $\mu$ 
have so far met with
little success \cite{LAT};
understandably, therefore, there has  been comparatively little parallel
development in relating correlation functions measured using
standard lattice field theory 
techniques to the phenomenology of dense systems. 
Exceptions are recent attempts to measure diquark condensates 
$\langle qq\rangle$ in Two Color QCD
(TCQCD) and Nambu -- Jona-Lasinio (NJL) models (see
eg. \cite{Hands,HW}) 
inspired by the
possibility that the QCD ground state at large $\mu$ and small $T$ is a color
superconductor \cite{RWA}. However, conceptually simpler issues such as the 
restoration of chiral symmetry and the spectrum of excitations
in a dense interacting medium continue to
present interesting
challenges. One obvious example is whether and how 
masses and widths of mesonic bound states such as the $\rho$ meson are modified
in a dense medium \cite{BRHL}. A broader question is how a Fermi
surface manifests itself in Euclidean simulations. Recall that color
superconductivity phenomenology is based on the BCS mechanism for electronic
superconductivity, which in turn is based on a small, 
though non-perturbative, deformation of a sharp Fermi surface.

As part of the learning process we have studied the simplest non-trivial 
model simulable with $\mu\not=0$ using standard algorithms, namely
the Gross-Neveu (GN) model in 2+1 dimensions.
Its Lagrangian density in Euclidean metric 
is written in terms of $4N_f$-component spinors $\psi,
\bar\psi$ as
\begin{equation}
{\cal L}=\bar\psi(\partial{\!\!\!/\,}+m)\psi-{g^2\over{2N_f}}(\bar\psi\psi)^2.
\end{equation}
In the chiral limit $m=0$ the model has a global Z$_2$ symmetry
$\psi\mapsto\gamma_5\psi$, $\bar\psi\mapsto-\bar\psi\gamma_5$.
At tree level the model's content is $N_f$ fermion flavors each with
parity-invariant mass $m$, interacting via a four-fermion contact term.
However at leading order in an expansion in $1/N_f$ it can be shown that for
sufficiently large coupling strength $g^2$ the
Z$_2$ symmetry is spontaneously broken by a condensate
$\langle\bar\psi\psi\rangle\not=0$ leading to a dynamically generated
fermion
mass gap $\Sigma_0=g^2\langle\bar\psi\psi\rangle\gg m$. The critical coupling 
$g_c^2$ at which the gap $\Sigma_0/\Lambda_{UV}\to0$
defines an ultra-violet stable fixed point of
the renormalisation group at which an interacting continuum limit may be taken.
This picture has been verified both at next-to-leading order in $1/N_f$
and by Monte
Carlo simulations with finite $N_f$ (eg. \cite{HKK1}).

The thermodynamic phase structure of the GN model is also known to leading order
in $1/N_f$ \cite{KRWP,HKK2}. At $T=0$, 
$\langle\bar\psi\psi\rangle$ remains constant as $\mu$ is increased until 
the Z$_2$ symmetry is restored in a first-order transition
at a critical $\mu_c=\Sigma_0$. At the same point the baryon density
$n\equiv N_f^{-1}\langle\bar\psi\gamma_0\psi\rangle$ 
jumps from zero to $\mu_c^2/2\pi$ and then
continues to rise quadratically as the Fermi disk 
of relativistic ``quark matter'' grows. This point is hence often
called the {\em onset\/}. Simulations with finite $N_f$ 
have again verified this picture with \cite{HKK2}, 
and importantly have showed that 
the first order nature of the chiral transition at $\mu_c\lapprox\Sigma_0$
persists at small but non-zero
$T$, implying the existence of a tricritical point in the $(\mu,T)$ plane
\cite{KS}. With currently available precision there is however no evidence for
a ``nuclear matter'' phase with both $\langle\bar\psi\psi\rangle\not=0,n\not=0$.

The high-$\mu$ phase of a related model, the 2+1$d$ NJL model, which has
a global SU(2)$_L\otimes$SU(2)$_R$ chiral symmetry, has also been extensively
studied \cite{HLM}. The main issue of interest has been whether diquark
condensation takes place at the Fermi surface in the isopsin-symmetric
degenerate matter formed for $\mu>\mu_c$, leading to a spontaneous
breakdown of baryon number symmetry whose physical manifestation is
superfluidity. Diquark pairing is favoured in this model
because the formulation with staggered lattice
fermions admits a scalar isoscalar $qq$ operator localised on a single 
lattice site \cite{HM}. It should be noted that in the current Z$_2$ model
a local $O(N_f)$ invariant diquark operator is forbidden by the
Pauli exclusion principle.  Although we have not studied the diquark sector in
the current work, we will see in Sec.~\ref{sec:ferm_disp} below
that the two models show dramatic differences in the spectrum of
spin-${1\over2}$ excitations around the Fermi surface.

In Sec.~\ref{sec:aux} we will use the technology of the $1/N_f$ expansion
in the so-called ``hard dense loop'' approach to calculate the long-range 
inter-particle interaction in medium, ie. for $\mu>\mu_c$. 
The main result is that the Debye mass
is infinite, but the plasma frequency $M_\sigma$ is finite and actually vanishes
as the onset $\mu\to\mu_{c+}$ is approached.
The main new feature emerging for $\mu>\mu_c$ is the existence of 
a new physical scale, the Fermi momentum $k_F$. Since the lowest energy
excitations of the ground state have $\vert\vec k\vert\approx k_F$, 
measurements of Euclidean
timeslice correlators with $\vec k\not=0$ are mandatory. The resulting energy
$E(\vec k)$ is known as the {\em dispersion relation\/}. The main numerical
results of the paper are thus 
dispersion relations in various channels of interest.
In the spin-${1\over2}$ fermion channel, 
the low-energy excitations about the Fermi surface, which may be
hole- or particle-like, are generically known as {\em quasiparticles\/}.
There is a well-developed framework for describing quasiparticles and their
interactions known as Fermi liquid theory. In Sec.~\ref{sec:fermiliquid} we will
present 
theoretical predictions of Fermi liquid properties to leading non-trivial
order in $1/N_f$. In Sec.~\ref{sec:friedel} we will describe another consequence
of the new scale; the wavefunction describing inter-particle spatial
correlations, which decays monotonically in the vacuum, now oscillates with
spatial frequency related to $k_F$.

Sec.~\ref{sec:num} describes the results of our numerical study. As mentioned
above, the fermion dispersion relation is presented in Sec.~\ref{sec:ferm_disp},
and mesonic (ie. $q\bar q$) channels are similarly investigated 
in Secs.~\ref{sec:sigma_disp}
and \ref{sec:meson_disp}. We will find that the predictions of the 
$1/N_f$ approach are verified with reasonable accuracy, and that in-medium
effects are clearly visible. In most meson channels
the lightest excitations are particle-hole pairs with effectively
zero energy whose temporal decay is hence algebraic rather than exponential, but
which nonetheless show a non-trivial dependence on the ratio $\vert\vec
k\vert/\mu$.
An interesting
phenomenon not described within the $1/N_f$ approach is a collective excitation
of the whole system (ie. not a quasiparticle) corresponding to a distortion
in the shape of the Fermi surface which propagates as {\em zero sound\/}.
Tentative evidence for an excitation of this form is presented in
Sec.~\ref{sec:meson_disp}. Finally in Sec.~\ref{sec:friedel_num} we present
results for wavefunctions in mesonic channels which clearly show the
oscillatory behaviour described above. After a brief summary, various technical
details of the calculation of relevant correlators 
in free-field theory are given
in two appendices.

\section{Theoretical Results}
\label{sec:theory}
In most theoretical approaches to the GN model it is convenient to encode
interactions by the introduction of an auxiliary scalar field $\sigma$ related
to the fermions via the equation of motion
$\sigma={g\over{\surd N_f}}\bar\psi\psi$. Although at tree level $\sigma$ is
non-propagating, the leading order $1/N_f$ expansion at $\mu=0$
predicts that it
acquires dynamics through quantum corrections due to virtual $q\bar q$ pairs,
resulting in a propagator $D_\sigma(k^2)$ of the form \cite{HKK1}
\begin{equation}
D_\sigma(k^2)\propto\cases{(k^2+4\Sigma_0^2)^{-1},&$k\ll\Sigma_0$;\cr
                           (k^2)^{-{1\over2}},&$k\gg\Sigma_0$.\cr}
\label{eq:Dsigma}
\end{equation}
For soft momenta $\sigma$ thus resmbles an orthodox scalar boson formed as 
a weakly-bound $q\bar q$ state. In the Born approximation $\sigma$ exchange
generates an interaction between the fermions of range $O(\Sigma_0^{-1})$ and
strength $O(N_f^{-1})$.
By contrast, as $k\to\infty$ $D_\sigma$ is a harder
function of $k$, though still softer than the original tree-level prediction
$D_\sigma\propto k^0$. The crossover from IR to UV is encoded as a branch cut 
in the complex $k$ plane, which means that the decay of $D_\sigma$ in Euclidean
time is not well-described by a simple pole. In the next subsection we will
extend the leading order calculation of $D_\sigma$ to $\mu>\mu_c$, restricting
our attention to soft momenta $k\ll\mu$. Even with this restriction
the resulting dispersion relation is surprisingly complicated.
The calculation also gives access to the forward
scattering amplitude between quasiparticles in the medium, which in turn enables
contact with the relativistic theory of Fermi liquids \cite{BC}.
It is hence possible to find quantitative relations 
between the Fermi liquid parameters $k_F$,
$\varepsilon_F$ and $\beta_F$, respectively the Fermi momentum, energy and
velocity.

\subsection{The $\sigma$ Propagator in Medium}
\label{sec:aux}
Here we examine the auxiliary scalar propagator $D_\sigma$ in the presence
of a non-zero baryon density. To leading order in the $1/N_f$ expansion 
(eg. \cite{HKK1}), we have 
\begin{eqnarray}
D_\sigma^{-1}(k)&=&1-\Pi(k;\mu)\nonumber\\
&=&g^2\left[{1\over\Sigma_0}\int_{q,\mu=0}
\mbox{tr}{1\over{iq{\!\!\! /\,}+\Sigma_0}}
+\int_{q,\mu>\mu_c}\mbox{tr}{1\over{iq{\!\!\! /\,}+\mu\gamma_0}}
{1\over{i(q{\!\!\!/\,}-k{\!\!\!/\,})+\mu\gamma_0}}
\right],
\end{eqnarray}
where $\Pi$ is the virtual fermion -- anti-fermion vacuum polarisation bubble,
we have used the gap equation at $\mu=0$ to express $1/g^2$ in terms of the
zero density gap $\Sigma_0$, and assumed that the gap vanishes in the 
integral defining $\Pi$ for $\mu>\mu_c$. It is technically more convenient
to 
evaluate the integrals for temperature $T>0$ and for arbitrary spatial
dimension $d\in(1,3)$; 
after tracing over Dirac indices we find
\begin{eqnarray}
D_\sigma^{-1}(k_0,\vec k)=4g^2T\sum_n\int{{d^d\vec q}\over{(2\pi)^d}}
\Biggl\{{1\over{\omega_n^2+\vec q^{\,2}+\Sigma_0^2}}-
{1\over{(\omega_n-i\mu)^2+\vec q^{\,2}}}\nonumber\\
-{{k_0(\omega_n-k_0-i\mu)+\vec k.(\vec q-\vec k)}\over
{[(\omega_n-i\mu)^2+\vec q^{\,2}][(\omega_n-k_0-i\mu)^2
+(\vec q-\vec k)^2]}}\Biggr\},\label{eq:intT}
\end{eqnarray}
where the sum is over discrete Matsubara modes $\omega_n=(2n-1)\pi T$.

The first two terms of (\ref{eq:intT}) are individually divergent, 
but their sum is finite. In the limit $T\to0$ (see eg. \cite{HKK2}) their
contribution is
\begin{equation}
{{4g^2}\over{(4\pi)^{d\over2}\Gamma(\textstyle{d\over2})}}\biggl\{
\int_0^\Lambda dq {q^{d-1}\over\sqrt{q^2+\Sigma_0^2}}-\int_\mu^\Lambda
dqq^{d-2}\biggr\}={{4g^2(\mu^{d-1}-\mu_c^{d-1})}
\over{(4\pi)^{d\over2}\Gamma(\textstyle{d\over2})(d-1)}},
\label{eq:int_stat}
\end{equation}
the critical chemical potential in $d+1$ dimensions in the large-$N_f$
limit being 
\begin{equation}
\mu_c=\left({{(1-d)\Gamma(\textstyle{d\over2})\Gamma(\textstyle{1\over2}-
\textstyle{d\over2})}\over{2\surd\pi}}\right)^{1\over{d-1}}\Sigma_0.
\end{equation}

The frequency sum for the momentum-dependent term may be evaluated 
using the formul\ae (eg. \cite{LeBellac})
\begin{eqnarray}
&-T&\!\!\!\sum_n \Delta^+(\omega_n,E_1)\Delta^-(\omega-\omega_n,E_2)=
\sum_{s_i=\pm1}
{{1-f_+(s_1E_1)-f_-(s_2E_2)}\over{
4s_1s_2E_1E_2(i\omega-s_1E_1-s_2E_2)}};\\
&iT&\!\!\!\!\!\sum_n\omega_n\Delta^+(\omega_n,E_1)\Delta^-(\omega-\omega_n,E_2)=
\!\!\!\sum_{s_i=\pm1}\!\!{{(\mu\!-\!s_1E_1)
(1\!-\!f_+(s_1E_1)\!-\!f_-(s_2E_2))}
\over{4s_1s_2E_1E_2(i\omega-s_1E_1-s_2E_2)}},\nonumber
\end{eqnarray}
where $\Delta^\pm(\omega,E)\equiv((\omega\mp i\mu)^2+E^2)^{-1}$, the
Fermi-Dirac function $f_\pm(x)\equiv(\exp((x\mp\mu)/T)+1)^{-1}$, and $E_i$ is
taken positive throughout. Using the identity $1-f_\pm(x)-f_\mp(-x)=0$,
and noting that for $\mu>0$ 
$\lim_{T\to0}f_+(E)=1-\theta(E-\mu)$, $\lim_{T\to0}f_-(E)=0$, 
we find in this limit
\begin{eqnarray}
g^2\int{{d^d\vec q}\over{(2\pi)^d}}{1\over{E_1E_2}}\Biggl\{
[-k_0^2+&&\!\!\!\!\!\!\!\!\!\!\!\!\!\!
\vec k.(\vec q-{\textstyle{1\over2}}\vec k)
-iE_1k_0]\biggl(
{{\theta(E_1-\mu)}\over{ik_0-E_1-E_2}}+\nonumber\\
{{\theta(E_2-\mu)-\theta(E_1-\mu)}\over
{ik_0-E_1+E_2}}\biggr)
&-&[-k_0^2+\vec k.(\vec q-{\textstyle{1\over2}}\vec k)+
iE_1k_0]{{\theta(E_2-\mu)}\over
{ik_0+E_1+E_2}}\Biggr\}
\end{eqnarray}
where the spatial loop momentum has been rerouted by $\vec q\to\vec
q+{1\over2}\vec k$ so that $E_{1,2}=\vert\vec q\pm{1\over2}\vec k\vert$.
We now restrict our attention to 
$k_0,\vert\vec k\vert\ll\mu$, thus probing the inter-particle 
interaction for soft momentum transfer in the so-called
{\em hard dense loop\/} (HDL) approximation \footnote{For asymptotically
large $k\gg\mu,\Sigma_0$ the 
form of the propagator reverts to $D^{-1}_\sigma(k)\propto k^{d-1}$
\cite{HKK1}.}.
The first and third terms together yield
\begin{eqnarray}
{{4g^2}\over{(4\pi)^{d\over2}\Gamma(\textstyle{d\over2})}}
\int_0^{2\pi}{{d\theta}\over{2\pi}}\int_{\mu
-{{{\vert\vec k\vert}\over2}\cos\theta+\cdots}}
^\infty dq q^{d-2}{{k_0^2+\vert\vec k\vert^2+ik_0\vert\vec k\vert\cos\theta
+O({k^3\cos\theta/q,k^4/q^2})}
\over{k_0^2+4q^2}}\nonumber\\
={g^2\mu^{d-3}\over{(4\pi)^{d\over2}\Gamma(\textstyle{d\over2})}}
\Bigl({{k_0^2+\vert\vec k\vert^2}\over{3-d}}
+i{k_0\over4\mu}\vert\vec k\vert^2
+O(k^4/\mu^2)\Bigr)
\label{eq:int_mom}
\end{eqnarray}
The integrand of the second term is non-zero only in a crescent-shaped
region near the Fermi surface; we find
\begin{eqnarray}
g^2{{\vert\vec k\vert}\over{(4\pi)^{d\over2}
\Gamma({d\over2})}}\int_{-{\pi\over2}}^{\pi\over2}{{d\theta}\over{2\pi}}
{\cos\theta\over{k_0^2+\vert\vec k\vert^2\cos^2\theta}}
\int^{\mu+{{\vert\vec k\vert}\over2}\cos\theta+\cdots}_{\mu-{{\vert\vec
k\vert}\over2}\cos\theta+\cdots}dqq^{d-3}\nonumber\;\;\;\;\;\;\;\;\;\;\;\;\;
\;\;\;\;\;\;\;\;\;\\
\times\Bigl(-2(k_0^2+\vert\vec k\vert^2)
+i{k_0\over q}\vert\vec k\vert^2\cos\theta\sin^2\theta+O(k^4/q^2)\Bigr)
&=& \label{eq:int_mom2}
\\
{{g^2\mu^{d-3}}\over{(4\pi)^{d\over2}\Gamma({d\over2})}}
{{\vert\vec k\vert^2}\over{k_0+(k_0^2+\vert\vec k\vert^2)^{1\over2}}}
\Biggl[-(k_0^2+\vert\vec k\vert^2)^{1\over2}+
i{{k_0\vert\vec k\vert^2}\over{4\mu[k_0+(k_0^2+\vert\vec k\vert^2)^{1\over2}]}}
&+&O\Bigl({k^3\over\mu^2}\Bigr)
\Biggr]\nonumber
\end{eqnarray}
Eqns.
(\ref{eq:int_stat}), (\ref{eq:int_mom}) and (\ref{eq:int_mom2})
combined yield
\begin{eqnarray}
D_\sigma^{-1}(k_0,\vec k)=
{{g^2\mu^{d-3}}\over{(4\pi)^{d\over2}\Gamma(\textstyle{d\over2})(3-d)}}
\Biggl[4{{(3-d)}\over{(d-1)}}\mu^{3-d}(\mu^{d-1}-\mu_c^{d-1})+\;\;\;\;\;\;\;
\;\;\;\;\;\;\;\;\;\;\;\;\;\;\;\;\;\;\;\;\;\;\label{eq_Dsigma}\\
k_0^2+\vert\vec k\vert^2
\biggl(1-{{(3-d)(k_0^2+\vert\vec k\vert^2)^{1\over2}}\over
{k_0+(k_0^2+\vert\vec
k\vert^2)^{1\over2}}}\biggr)
+i{{k_0\vert\vec k\vert^2}\over{4\mu}}
\biggl(1+{{\vert\vec k\vert^2}\over
{[k_0+(k_0^2+\vert\vec
k\vert^2)^{1\over2}]^2}}\biggr)
+O\Bigl({k^4\over\mu^2}\Bigr)\Biggr].\nonumber
\end{eqnarray}

The first observation to make 
is that in the static limit $k_0=0$ the momentum dependence
of (\ref{eq_Dsigma}) vanishes for the physical value $d=2$, implying 
complete screening of the static potential due to $\sigma$ exchange, or in other
words infinite Debye mass.
By contrast, 
for $\vec k=\vec0$ $D_\sigma$ is proportional to a conventional zero-momentum
boson propagator
with mass 
\begin{equation}
M^2_\sigma=4{(3-d)\over(d-1)}\mu^{3-d}(\mu^{d-1}-\mu_c^{d-1})
\end{equation}
or for the physical $d=2$
\begin{equation}
M_\sigma=2\sqrt{\mu(\mu-\mu_c)}.
\label{eq:msigma}
\end{equation}
Just above the transition, therefore, there is
a tightly-bound state whose 
decay at rest 
into a particle - anti-particle pair is Pauli-blocked.
The free-field result $M_\sigma=2\mu$ is only approached as $\mu\to\infty$.
$M_\sigma$ may also be interpreted as the {\em plasma
frequency\/} of the system.

\begin{figure}[htb]
\bigskip\bigskip
\begin{center}
\epsfig{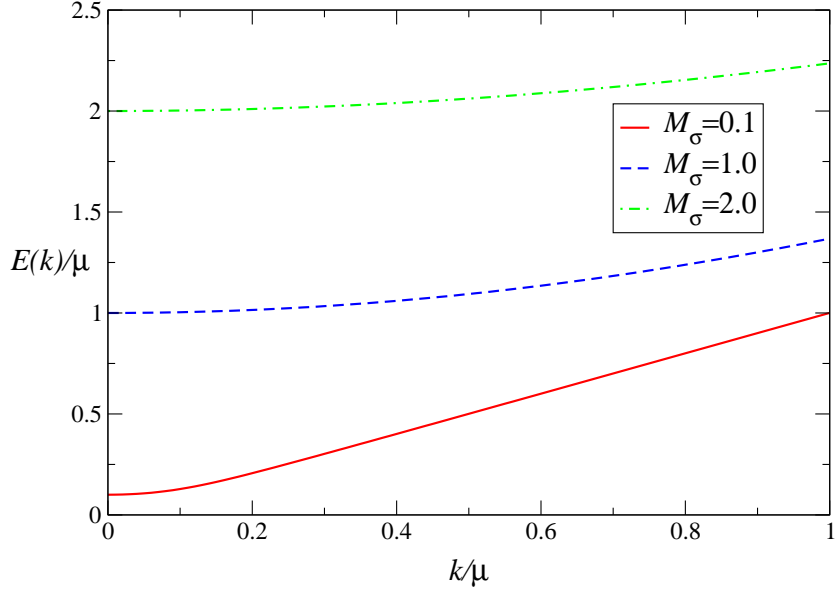}
\end{center}
\caption{The dispersion relation $E(\vert\vec k\vert)$ in units of $\mu$
for various values of $M_\sigma$.}
\label{fig:disp_sig}
\end{figure}
For states in 
motion, things are more complicated. For $\vert\vec k\vert\ll\vert
k_0\vert$ in $d=2$ we have
\begin{equation}
D^{-1}_\sigma(k_0,\vec k)\propto M_\sigma^2+k_0^2+{1\over2}\vert\vec k\vert^2
+i{{k_0\vert\vec k\vert^2}\over{4\mu}}+O\Bigl({{\vert\vec k\vert^4}\over k_0^2}
\Bigr)\label{eq_pole}.
\end{equation}
Eqn. (\ref{eq_pole}) yields two pure imaginary poles in the complex $k_0$ plane
leading to solutions 
with differing decay lengths
in forward and backward directions in Euclidean time. In principle this may be
probed by a Euclidean lattice simulation. However, linear response
theory dictates that on analytic continuation to Minkowski metric,
the physically relevant solution giving the
energy $E(\vec k)$ of a collective excitation of the system 
is given by poles of
the {\em retarded\/} propagator \cite{LeBellac}, which corresponds to the 
solution of (\ref{eq_pole}) with negative imaginary part:
\begin{equation}
E(\vec k)=\biggl(M_\sigma^2+{1\over2}\vert\vec k\vert^2\biggr)^{1\over2}
+{{\vert\vec
k\vert^2}\over{8\mu}}\simeq M_\sigma+\vert\vec k\vert^2
\biggl({1\over{4M_\sigma}}+{1\over{8\mu}}\biggr).
\end{equation}
In the limit $\mu\to\infty$ this becomes $E\simeq M_\sigma+\vert\vec
k\vert^2/2M_\sigma$, appropriate for a free non-relativistic particle of mass 
$M_\sigma$. Another limit we may consider is $\mu\to\mu_{c+}$, in which case
$M_\sigma^2\ll\vert\vec k\vert^2\simeq-k_0^2$ in
(\ref{eq_Dsigma}), yielding the
dispersion relation of a standard relativistic boson:
\begin{equation}
E^2(\vec k)\simeq M_\sigma^2+\vert\vec k\vert^2.
\end{equation}
Dispersion curves obtained from 
a numerical solution of (\ref{eq_Dsigma}) for three different $M_\sigma$ are
shown in Fig.~\ref{fig:disp_sig}.

\subsection{Fermi Liquid Properties}
\label{sec:fermiliquid}
Given what we know about the fermion - fermion interaction in medium from the
calculation of $D_\sigma$, it is 
possible to make quantitative statements in the context of the Fermi liquid
description of degenerate fermions, first developed for
non-relativistic systems such as $^3$He by Landau \cite{Landau} (see also
\cite{Landau2}). The generalisation to relativistic systems has subsequently
been given by Baym and Chin \cite{BC}. The essential physical idea is that
the dominant low-energy excitations in the neighbourhood of the Fermi surface
are long-lived quasiparticles having energy $\varepsilon_{\vec k}$,
width $\propto(\varepsilon_{\vec k}-\mu)^2$  and
equilibrium distribution $n_{\vec k}$ related by the Fermi-Dirac function
\begin{equation}
n_{\vec k}=f_+(\varepsilon_{\vec k}).
\end{equation}
For $T=0$
we expect $\varepsilon_{\vec k}$ to have the form
\begin{equation}
\varepsilon_{\vec k}\simeq\mu+\beta_F(\vert\vec k\vert-k_F)
\label{eq:FL}
\end{equation}
where $k_F$, $\beta_F\equiv\vert\vec\nabla_{\vec k}\,\varepsilon_{\vert\vec
k\vert=k_F}\vert$ 
are respectively the Fermi momentum and Fermi velocity,
which in the absence of further information are phenomenological parameters.
The quasiparticle energy is implicitly determined by the equation for its 
variation under a small departure from equilibrium;
\begin{equation}
\delta\varepsilon_{\vec k}=\int {{d^dk^\prime}\over{(2\pi)^d}}{\cal F}_
{\vec k,\vec
k^{\prime}}\delta n_{\vec k^{\prime}},
\end{equation}
where ${\cal F}_{\vec k,\vec k^{\prime}}={\cal F}_{\vec k^{\prime},\vec k}$
is the {\em Fermi liquid interaction\/}, 
which is in turn related to the two particle 
forward scattering amplitude 
\begin{equation}
{\cal F}_{\vec k,\sigma,\vec k^{\prime},\sigma^\prime}=
{1\over{4\varepsilon_{\vec
k}\varepsilon_{\vec k^{\prime}}}}{\cal M}_{\vec k,\sigma,\vec
k^{\prime},\sigma^\prime},
\end{equation} 
where we have retained explicit spin states $\sigma,\sigma^\prime$ and ${\cal
M}$ is the Lorentz invariant matrix element.
Landau's theory becomes predictive
if this amplitude is calculable in medium.

\begin{figure}[htb]
\bigskip\bigskip
\begin{center}
\epsfig{file=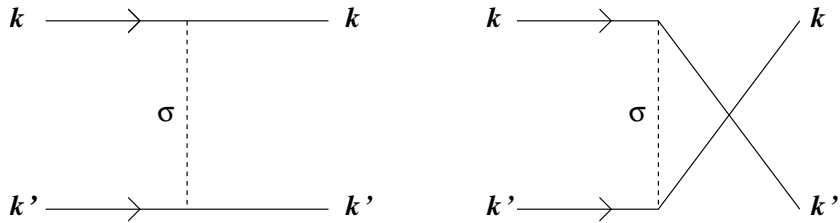, width=11cm}
\end{center}
\caption{Feynman diagrams for forward scattering, showing direct (left) and
exchange (right) contributions.}
\label{fig:Feynman}
\end{figure}
To leading non-trivial order in $1/N_f$ there are both direct and exchange
contributions to ${\cal M}$, as shown in Fig.~\ref{fig:Feynman}, 
but in the chiral
limit only the exchange term is non-vanishing. After averaging over spin states
we find \cite{BC}
\begin{equation}
{\cal F}_{\vec k,\vec k^{\prime}}={g^2\over{4N_f}}
\left[1-{{\vec k.\vec k^{\prime}}\over{\varepsilon_{\vec k}\varepsilon_{\vec
k^{\prime}}}}\right]D_\sigma(\varepsilon_{\vec k}-\varepsilon_{\vec k^
{\prime}},
\vec k-\vec k^{\prime}).
\end{equation}
Specialising to $d=2$ and using the result (\ref{eq_Dsigma}) for $D_\sigma$ 
we find that for 
$\vec k,\vec k^{\prime}$ on the Fermi surface separated by angle $\theta$,
\begin{equation}
{\cal F}_{\vec k,\vec k^{\prime}}=
{{\pi\mu}\over{N_fM_\sigma^2(\mu)}}(1-\cos\theta)
\label{eq:interaction}
\end{equation}
with $M_\sigma^2=4\mu(\mu-\mu_c)$.
This particularly simple form arises due to the vanishing spatial momentum 
dependence of 
$D_\sigma$ for zero energy exchange since the $\sigma$ has infinite 
Debye mass in medium. The factor of $N_f^{-1}$ ensures that ${\cal F}$ can
be treated systematically as a weak correction to the leading
order case of non-interacting particles; note that this would not hold
if direct interactions were involved, since that would result in a
compensating ``degeneracy factor'' of $N_f$ in the numerator.

It is possible to derive relations between the Fermi liquid parameters. For
instance, by requiring consistency of the above framework under Lorentz boosts
the following relation may be derived \cite{BC}:
\begin{equation}
\varepsilon_{\vec k}\vec\nabla_{\vec k}\,\varepsilon_{\vec k}=\vec k
+\int{{d^2k^\prime}\over{(2\pi)^2}}{\cal F}_{\vec k,\vec k^{\prime}}
\varepsilon_{\vec
k^{\prime}}\vec\nabla_{\vec k^{\prime}}n_{\vec k^{\prime}}.
\end{equation}
Dotting this equation with the unit vector $\hat{\vec k}$, and noting that 
at $T=0$, $n_{\vec k^{\prime}}=\theta(\mu-\varepsilon_{\vec k^{\prime}})$
and $\beta_F\delta(\varepsilon_{\vec k^{\prime}}-\mu)=\delta(\vert\vec
k^{\prime}\vert-k_F)$ we arrive at the following relation at the Fermi
surface:
\begin{equation}
\mu\beta_F+\mu k_F{\mathfrak g}{{\pi\mu}\over{N_fM_\sigma^2(\mu)}}
\int_0^{2\pi}{{d\theta}\over{4\pi^2}}
(1-\cos\theta)\cos\theta = k_F,
\end{equation}
ie.
\begin{equation}
\beta_F={k_F\over\mu}+{\mathfrak g}{{k_F\mu}\over{4N_fM_\sigma^2(\mu)}}.
\label{eq_betaF}
\end{equation}
Here ${\mathfrak g}$ 
is the degeneracy of the states participating in the exchange
interaction; for the Z$_2$ GN model ${\mathfrak g}=2$ counts the number of spin
states, whereas 
${\mathfrak g}=4$ for the
isospin-symmetric matter described by the NJL model
when $\pi$  exchange is also taken into
account.

Similarly, the compressibility follows from differentiating the relation
$\mu=\varepsilon_{k_F}$ with respect to $n$:
\begin{equation}
{{\partial\mu}\over{\partial n}}={{\partial\varepsilon_k}\over{\partial
k_F}}
{{\partial k_F}\over{\partial n}}
+\bar{\cal F} 
=\beta_F{{2\pi}\over{{\mathfrak g}k_F}}+{{\pi\mu}\over{N_fM_\sigma^2(\mu)}},
\label{eq_comp}
\end{equation}
where we have used the equality of particle and quasiparticle densities
$n=\int 
{{d^2\vec k}\over(2\pi)^2} n_{\vec k}$ and the bar denotes the average over the
circle. Using relation (\ref{eq_betaF}) we then derive the {\em first
sound\/} velocity
$\beta_1$:
\begin{equation}
\beta_1^2\equiv{n\over\mu}{{\partial\mu}\over{\partial n}}=
{k_F^2\over{2\mu^2}}+{{3{\mathfrak g}k_F^2}\over{8N_fM_\sigma^2(\mu)}}.
\end{equation}
To find the relation between $\varepsilon_{k_F}=\mu$ and $k_F=(4\pi n/{\cal
G})^{1\over2}$ we need to integrate (\ref{eq_comp}):
\begin{equation}
\int_{k_{F_c}}^{k_F}k_Fdk_F=\int_{\mu_c}^\mu{{\mu d\mu}\over
{1+{{3{\mathfrak g}}\over{16N_f}}{\mu\over{\mu-\mu_c}}}}.
\label{eq_int}
\end{equation}
The integration limits specify the conditions at onset; note that
$k_{F_c}=\mu_c+O(N_f^{-1})=\Sigma_0+O(N_f^{-1})$ 
with the corrections given by the two-loop gap
equation. The solution of (\ref{eq_int}) 
expressed consistently to $O(N_f^{-1})$ is
\begin{equation}
k_F^2=\mu^2+(k_{F_c}^2-\mu_c^2)-{{3{\mathfrak g}}\over{16N_f}}\left[
\mu^2-\mu_c^2+2\mu_c(\mu-\mu_c)+2\mu_c^2\ln\left({{\mu-\mu_c}\over{3{\cal
G}\mu_c/16N_f}}\right)\right].
\label{eq_solint}
\end{equation}
We can combine this with (\ref{eq_betaF}) to get an absolute prediction for 
$\beta_F$. For $\mu\gg\mu_c$ we find
\begin{equation}
{k_F\over\mu}=1-{{3{\mathfrak g}}\over{32N_f}};\;\;
\beta_F=1-{{\mathfrak g}\over{32N_f}},
\label{eq:1/N}
\end{equation}
which, happily, is consistent with the causality requirement $\beta_F\leq1$.
In fact, the full solution (\ref{eq_solint}) violates
this condition for $\mu-\mu_c\approx O(\mu_c)$, although a precise
statement is impossible without knowledge of $k_{F_c}$ to $O(N_f^{-1})$.
This suggests that the HDL approximation leading to
(\ref{eq_Dsigma}) is insufficient at such small $\mu$.

\subsection{Friedel Oscillations}
\label{sec:friedel}

Another characteristic of a Fermi surface can be exposed by considering 
spatial correlations between $q\bar q$ pairs. A convenient way to do this is to
consider correlation functions of the form
\begin{eqnarray}
C(\vec y;x_0)&=&\sum_{\vec x}\langle \bar\psi\Gamma\psi(\vec0,0)\;\;
\bar\psi\Gamma\psi(\vec x+\vec y,x_0)\rangle\\
&=&\mbox{tr}\int_p\int_q\Gamma{{e^{ipx}}\over{ip{\!\!\!/\,}+\mu\gamma_0+M}}
\Gamma{{e^{-iqx}e^{-i\vec q.\vec y}}\over{iq{\!\!\!/\,}+\mu\gamma_0+M}}
\end{eqnarray}
where in $2+1$ dimensions $\int_p=\int d^3p/(2\pi)^3$ 
and the matrix $\Gamma$ determines the quantum numbers of the
channel. $M$ is the fermion mass, which we will take to have the 
leading order form $M(\mu)=\Sigma_0\theta(\mu_c-\mu)$. Performing
the Dirac trace and the sum over $\vec x$ yields
\begin{equation}
C(\vec y;x_0)={4\over(2\pi)^{4}}\int\! dq_0\int\! dp_0\int\! d^2\vec p
{{(\pm(p_0-i\mu)(q_0-i\mu)\pm\vec p^{\,2}+M^2)
e^{i(p_0-q_0)x_0}e^{-i\vec p.\vec y}
}\over{[(p_0-i\mu)^2+\vec
p^{\,2}+M^2][(q_0-i\mu)^2+\vec p^{\,2}+M^2]}}
\label{eq:wvfn}
\end{equation}
where the $\pm$ signs correspond to pseudoscalar and scalar channels
respectively. 

First consider the chirally broken phase $\mu<\mu_c=\Sigma_0$; 
the $p_0$ ($q_0$) integral may be performed
using Cauchy's theorem by completing the contour in the upper (lower)
half-plane,
picking up residues at the poles at $i\mu\pm i\sqrt{\vec p^{\,2}+M^2}$.
Therefore 
\begin{eqnarray}
C(\vec y;x_0)&=&{2\over(2\pi)^2}\int d^2\vec p e^{-2x_0\sqrt{\vec p^{\,2}+M^2}}
e^{-i\vec p.\vec y}\times\cases{1,& pseudoscalar\cr
     -{{\vec p^{\,2}}\over{\vec p^{\,2}+M^2}},&scalar\cr}\\
&=&2\int_0^\infty pdpJ_0(py)e^{-2x_0\sqrt{p^2+M^2}}.
\label{eq:wvfn2}
\end{eqnarray}
where $J$ is the Bessel function and in the second line we have specialised to
the pseudoscalar case. For large $x_0$, the final integral over $p$ can now be
approximated using Laplace's method of asymptotic expansion, by expanding
the integrand in powers of $p/M$, yielding
\begin{equation}
\lim_{x_0\to\infty}C(\vec y;x_0)\sim {M\over x_0}e^{-2Mx_0}\exp\left(-
{{\vert\vec y\vert^2M}\over4x_0}\right).
\end{equation}
The general profile of the wavefunction is a gaussian with width increasing as
$\surd x_0$. 

In the chirally restored phase $\mu>\mu_c$, $M$ vanishes. For $\vert\vec
p\vert<\mu$, both poles of the integrand of (\ref{eq:wvfn}) lie in the upper
half-plane, making the $q_0$ integral vanish. The effect is to raise the lower
limit of the integral of (\ref{eq:wvfn2}) from 0 to $\mu$. The asymptotic
expansion is now made by expanding the integrand in powers of $({p\over\mu}-1)$,
with the resulting form
\begin{equation}
\lim_{x_0\to0}C(\vec y;x_0)\sim{\mu\over x_0}e^{-2\mu x_0}J_0(\mu\vert\vec
y\vert).
\label{eq:friedel}
\end{equation}
The wavefunction profile no longer changes with $x_0$, but instead oscillates
with a spatial frequency determined by $\mu$, which to this order may be
identified with the Fermi momentum $k_F$. 

The oscillations observed in $C(\vec y;x_0)$ are characteristic
of a sharp Fermi surface and are reminiscent of oscillations in
either the density-density correlation function, or the screened inter-particle
potential, in degenerate systems generically known as {\em Friedel
Oscillations\/} \cite{FW}. The resemblance is only qualitative, however,
because while the large-$y$ form of (\ref{eq:friedel}) is $C\propto
y^{-{1\over2}}\sin(k_Fy-{\pi\over4})$, the form of eg. the screened potential
in high-density QED is $y^{-3}\cos(2k_Fy)$, and even 
$y^{-4}\sin(2k_Fy)$ in the relativistic limit $m_e\ll\mu$ \cite{KT}.
The disparity is due in part to the lower dimensionality, and partly because 
the function $C(\vec y;x_0)$ probes correlations between fermion field variables
rather than scalar densities. In Appendix A we present a similar asymptotic
analysis of the density-density correlator 
$C_{nn}(\vec y;x_0)=\sum_{\vec x}\langle j_0(\vec0,0)j_0^\dagger(\vec x,x_0)
j_0(\vec0,0)j_0^\dagger(\vec x+\vec y,x_0)\rangle$, where
$j_0(x)$ is the local baryon density $\bar\psi\gamma_0\psi(x)$, and find
a decay with $y$ more closely resembling the forms in the literature. In
particular, because $C_{nn}$ describes correlations between particle-hole pairs at the
Fermi surface, the asymptotic spatial frequency of the oscillations is
$2k_F$ rather than $k_F$. Because of the similar origin, however, we will 
refer to oscillations of the form (\ref{eq:friedel}) as Friedel oscillations in
what follows.

\section{Numerical Results}
\label{sec:num}

In this section we present results from a numerical simulation of the 
Z$_2$ GN model using staggered lattice fermions. 
The lattice formulation and simulation algorithm
are described in
detail in Refs.~\cite{HKK1,HKK2}. We used a $32^2\times48$ lattice 
and throughout chose a bare fermion mass $m=0$ and
$N=2$ lattice flavors. Unless
otherwise stated the coupling constant used was $a/g^2=0.75$, corresponding to 
a physical fermion mass at $\mu=0$ of $\Sigma_0a\simeq0.17$. At each value 
of $\mu$ between 2000 and 4000 hybrid Monte Carlo trajectories of mean length
1.0 were taken. All results have
been taken in the chirally-restored phase with $\mu>\mu_c$, whose value is
estimated to be $\mu_ca\simeq0.16(1)$. Values for the baryon density per
staggered fermion flavor $na^2$ 
for values of chemical potential $\mu a\in[0.2,0.8]$ are given in
Table~\ref{tab:n}, showing that the density rises steadily as $\mu$ is
increased. The reported densities are roughly twice the free-field value
$n=\mu^2/2\pi$, consistent with fermion doubling; $N$ flavors of staggered
lattice fermion describe $N_f=2N$ continuum flavors in 2+1 dimensions, so that
$N_f=4$ in our simulation.
Henceforth we will quote results in units in which the lattice spacing 
$a=1$.
\begin{table}[h]
\centering
\caption{Baryon density $n=N^{-1}\langle\bar\psi\gamma_0\psi\rangle$}
\smallskip
\label{tab:n}
\setlength{\tabcolsep}{1.5pc}
\begin{tabular}{|ll|}
\hline
$\mu$ & $n$   \\
\hline
0.2     & 0.0113(1)   \\
0.4  &  0.0520(1)   \\
0.5  &  0.0865(1)   \\
0.6  &  0.1394(1)  \\
0.8  &  0.3127(1)  \\
\hline
\end{tabular}
\end{table}

\subsection{Fermion Dispersion Relation}
\label{sec:ferm_disp}

\begin{figure}[htb]
\bigskip\bigskip
\begin{center}
\epsfig{file=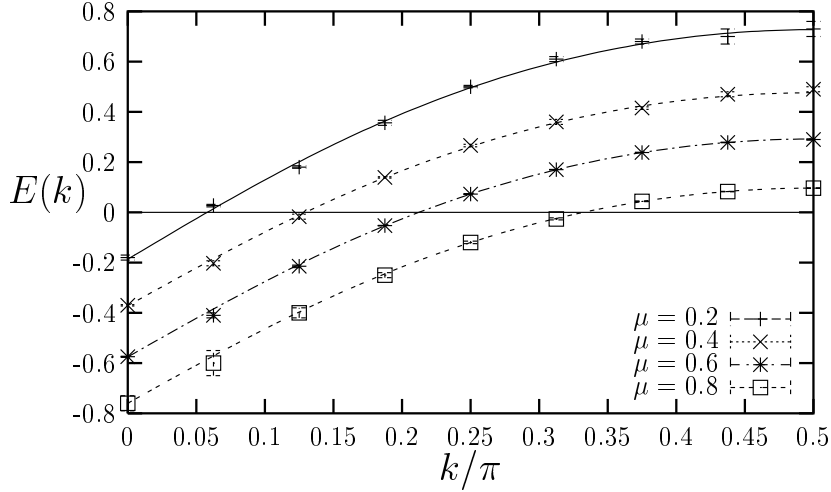, width=11cm}
\end{center}
\caption{The fermion dispersion relation $E(\vert\vec k\vert)$
at four values of $\mu$.}
\label{fig:ferm_disp}
\end{figure}

The fermion timeslice propagator at fixed spatial momentum $\vec k$ is defined
by ${\cal G}(\vec k,x_0)=\sum_{\vec x}{\cal G}(\vec 0,0;\vec x,x_0)e^{-i\vec
k.\vec x}$, where ${\cal G}(x,y)\equiv\langle\psi(x)\bar\psi(y)\rangle$.
The decay in Euclidean time $x_0$ then yields the fermion dispersion relation
$E(\vec k)$. For a degenerate system with Fermi momentum $k_F$, excitations 
with $\vert\vec k\vert<k_F$ 
are hole-like and those with $\vert\vec k\vert>k_F$ particle-like; the lowest 
energy excitations are those in the neighbourhood of the Fermi surface with 
$\vert\vec k\vert\simeq k_F$. 
Fig.~\ref{fig:ferm_disp} shows $E(\vec k)$ measured for $\vec k$
parallel to the $x$-axis for
four values of $\mu>\mu_c$, using the procedure specified in \cite{HLM}. 
In order to obtain a smooth curve we plot $E(\vert \vec k\vert)$ as negative for
$\vert\vec k\vert<k_F$. 
Note also that $\vert\vec k\vert$ is plotted as a
fraction of $\pi$ -- since the staggered fermion action is only 
invariant under translations of length $2a$ the maximum momentum
which can be probed is $\pi/2$.

It is obvious that there is no mass gap at the Fermi surface characteristic
of a BCS instability, since
for certain combinations of $\mu$ and $\vec k$ our data have 
$E(\vert\vec k\vert)\approx 0$.
Indeed, it is possible to fit a smooth function 
\begin{equation}
E(\vert\vec k\vert)=-E_0+D\sinh^{-1}(\sin\vert\vec k\vert)
\end{equation}
to the data, which corrects the Fermi liquid form (\ref{eq:FL}) for 
lattice discretisation effects, and permits the extraction of an effective
Fermi momentum $K_F$ and velocity $\beta_F$ given by \cite{HLM}
\begin{equation}
K_F=\sinh^{-1}(\sin k_F)={E_0\over D};\;\;\;
\beta_F=D{{\cosh E_0}\over{\cosh K_F}}.
\end{equation}

\begin{table}[h]
\centering
\caption{The Fermi momentum and Fermi velocity.}
\smallskip
\label{tab:FL}
\setlength{\tabcolsep}{1.5pc}
\begin{tabular}{|llll|}
\hline
$\mu$ & $K_F$ &  $\beta_F$  & $K_F/\mu\beta_F$ \\
\hline
0.2     & 0.18(1) & 1.04(2) & 0.87(5)  \\
0.4  & 0.386(6) & 0.96(1)   & 1.00(2) \\
0.6  & 0.584(3) & 0.98(1)   & 0.99(1) \\
0.8  & 0.782(14) & 0.96(1)  & 1.02(2) \\
\hline
\end{tabular}
\end{table}
The fitted values of $K_F,\beta_F$ are given in Table~\ref{tab:FL}.
The departures from free-field values are small and of similar magnitude
to the predictions (\ref{eq:1/N})
of the $1/N_f$ expansion
$K_F/\mu\simeq0.953$, $\beta_F\simeq0.984$.
Also shown is the ratio $K_F/\mu\beta_F$, which apart from $\mu=0.2$ shows
no significant deviation from its free-field value 1.
This constrasts sharply with 
Monte Carlo studies of the 2+1$d$ NJL model, for which ${\mathfrak g}=4$,
where values in the range 1.3 - 1.5 were observed \cite{HLM}. 
This suggests that the quasiparticle
spectrum observed in that model is determined by physics outside the scope of
the $1/N_f$ expansion, and as such the current result provides 
indirect support for the exotic gapless superfluid scenario 
proposed in
\cite{HLM}.

\subsection{The $\sigma$ Dispersion Relation}
\label{sec:sigma_disp}

Just as in the continuum $1/N_f$ expansion, the lattice formulation of the
GN model encodes the interactions between the fermions via an auxiliary bosonic
scalar field $\sigma$ whose value is given by the equation of motion
$\sigma={g\over{\surd N_f}}\bar\psi\psi$ (in this paper we define $\sigma$
with dimension ${3\over2}$). For $\mu=0$ and $k\ll\Sigma_0$
the auxiliary propagator (\ref{eq:Dsigma}) resembles
that of a standard boson of mass $2\Sigma_0$. 
Since this mass coincides with the
threshold for an $q\bar q$ continuum, $D_\sigma$ is dominated
by a branch cut rather than an isolated pole in the complex $k$-plane
making standard lattice spectroscopy techniques difficult to apply \cite{HKK1}.
A recent study using the Maximum Entropy Method has shown evidence for
a non-zero binding energy in the $\sigma$ channel for finite $N_f$ \cite{MEM}.

\begin{figure}[]
\bigskip\bigskip
\begin{center}
\epsfig{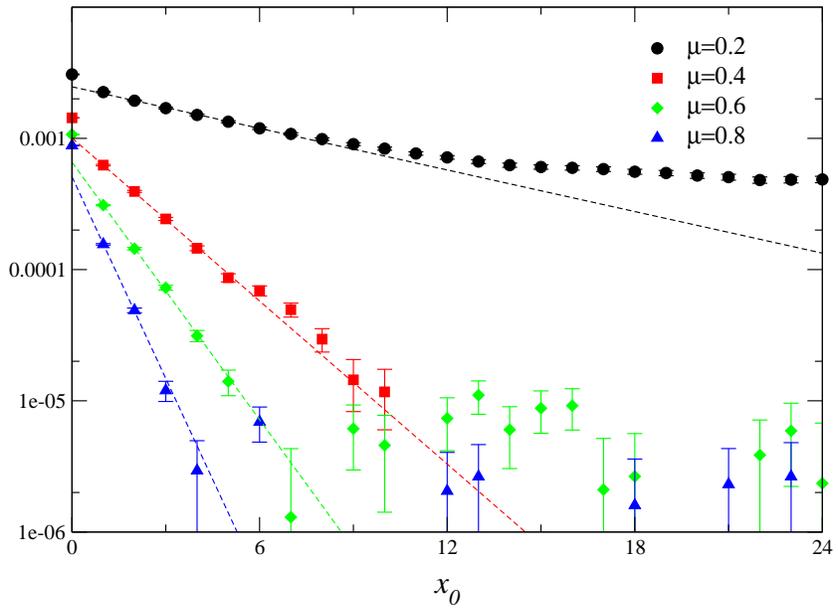}
\end{center}
\caption{The $\sigma$ timeslice correlator ${\cal D}(x_0)$ at $\vec k=\vec 0$
for 4 values of $\mu$.}
\label{fig:sigma_corr}
\end{figure}

\begin{figure}[]
\bigskip\bigskip
\begin{center}
\epsfig{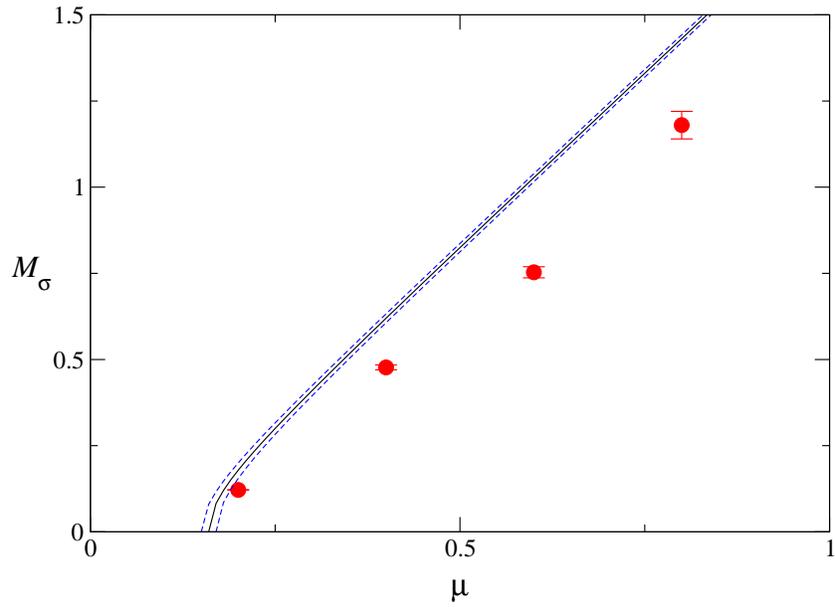}
\end{center}
\caption{$M_\sigma$ as a function of $\mu$. The line is the
leading order $1/N_f$ prediction (\ref{eq:msigma}).}
\label{fig:msigma}
\end{figure}

For $\mu>\mu_c$, however, the calculation of Sec.~\ref{sec:aux} suggests
that the $\sigma$ is more tightly bound. We have studied this by measuring the
auxiliary timeslice propagator ${\cal D}(\vec k,\vert x_0\vert)=
\sum_{\vec x,x_0=\pm\vert x_0\vert}\langle\sigma(\vec
0,0)\sigma(\vec x,x_0)\rangle e^{-i\vec k.\vec x}$, once again
choosing $\vec k$ along the $x$-axis\footnote{Note that as
discussed below eqn.~(\ref{eq_pole}) ${\cal D}$ 
may not in fact be time-reversal symmetric for $\vec k\not=\vec 0$.}.
Fig.~\ref{fig:sigma_corr} plots ${\cal D}(x_0)$ for $\vec k=\vec 0$, showing
that with the exception of $\mu=0.2$, simple pole fits of the form
${\cal D}(x_0)\propto e^{-M_\sigma x_0}$ 
describe the data reasonably well
before it descends into noise. Values of $M_\sigma$ extracted from fits 
to timeslices $x_0\in[1,5]$ are plotted in Fig.~\ref{fig:msigma}. Also shown is
the leading order prediction $M_\sigma=2\sqrt{\mu(\mu-\mu_c)}$, with a value
0.16(1) assumed for $\mu_c$. The numerical results fall 20-30\% lower than 
the leading order prediction, which is consistent with a correction of 
$O(N_f^{-1})$.

\begin{figure}[]
\bigskip\bigskip
\begin{center}
\epsfig{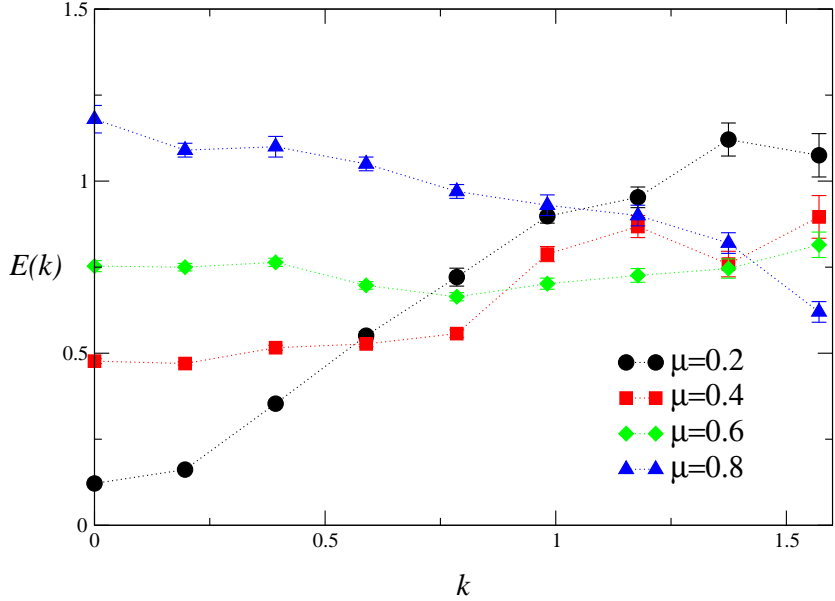}
\end{center}
\caption{The $\sigma$ dispersion relation $E(\vert\vec k\vert)$ 
for 4 values of $\mu$.}
\label{fig:sigdisp}
\end{figure}

Results from simple pole fits to ${\cal D}(\vec k,\vert x_0\vert)$ 
for $\vert\vec
k\vert\leq{\pi\over2}$
yielding the $\sigma$ dispersion relation $E(k)$ are plotted in
Fig.~\ref{fig:sigdisp}
\footnote{Since under the global Z$_2$ 
$\sigma(x)\mapsto(-1)^{x_0+x_1+x_2}\sigma(x)$,  in the symmetric phase 
the propagator obeys ${\cal D}(k_i)={\cal D}(\pi-k_i)$.}.
Only for the lowest value $\mu=0.2$ is there any marked
change in $E$ as $k$ is increased, in which case $E\approx k$. For larger $\mu$
the dispersion is flatter. There is qualitative agreement with the 
leading order $1/N_f$ result shown in
Fig.~\ref{fig:disp_sig}, although an improved understanding of discretisation
effects close to $\vert\vec k\vert={\pi\over2}$ would 
be needed to make the comparison
more quantitative.
Nonetheless, Figs.~\ref{fig:sigma_corr} 
and \ref{fig:sigdisp} demonstrate unambiguously that in-medium
effects are observed in the lattice simulation.

\subsection{Mesonic Dispersion Relations and Zero Sound}
\label{sec:meson_disp}

Next we investigate mesonic (ie. $q\bar q$) states.
We have studied mesonic correlation functions
${\cal C}_\Gamma(\vec k,x_0)=\sum_{\vec x}\langle j_\Gamma(\vec 0,0)
j_\Gamma^\dagger(\vec
x,x_0)\rangle e^{-i\vec k.\vec x}$ where the bilinears $j_\Gamma(x)$ 
are defined with
scalar, pseudoscalar or vector quantum numbers;
in terms of staggered fermion fields $\chi,\bar\chi$ the operators are 
\begin{equation}
j_{\One}(x)=\bar\chi_x\chi_x;\;\; 
j_{\gamma_5}(x)=\varepsilon_x\bar\chi_x\chi_x;\;\;
j_{\gamma_i}(x)={\eta_{ix}\over2}[\bar\chi_x\chi_{x+\hat\imath}+
                          \bar\chi_{x+\hat\imath}\chi_x],
\end{equation}
where $\eta_{1x}=(-1)^{x_0}$, $\eta_{2x}=(-1)^{x_0+x_1}$ and
$\varepsilon_x=(-1)^{x_0+x_1+x_2}$. As before we study the timeslice
correlators in each channel as a function of spatial momentum $\vec
k\parallel\hat x$. 

\begin{figure}[]
\bigskip\bigskip
\begin{center}
\epsfig{file=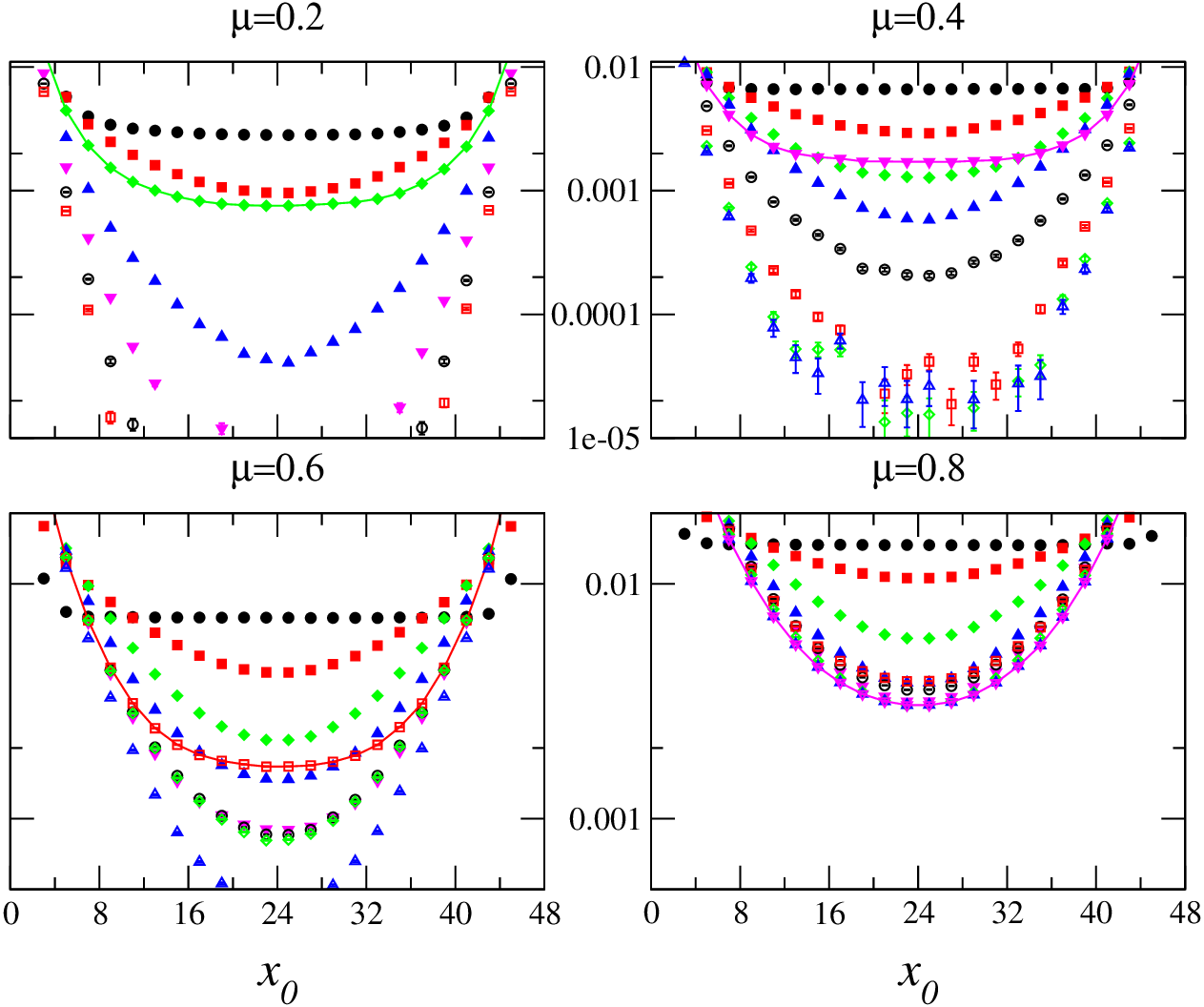, width=13cm}
\end{center}
\caption{Pseudoscalar correlator ${\cal C}_{\gamma_5}(\vert\vec k\vert,x_0)$ 
at 4 different $\mu$
for momenta $\vert\vec k\vert=0$ (filled circles),
${\pi\over16}$ (filled squares), ${\pi\over8}$ (filled diamonds),
${3\pi\over16}$ (filled up triangles), ${\pi\over4}$ (filled down triangles),
${5\pi\over16}$ (open circles), ${3\pi\over8}$ (open squares), ${7\pi\over16}$
(open diamonds) and  ${\pi\over2}$ (open up triangles).}
\label{fig:psdisperse}
\end{figure}

As expected for $\mu>\mu_c$ where the Z$_2$ global symmetry 
is restored, the results in scalar and pseudoscalar channels coincide. The 
results on odd timeslices, averaged over $k$ and $\pi-k$, are shown for four
different $\mu>\mu_c$ in Fig.~\ref{fig:psdisperse}. Note the difference in 
the vertical logarithmic scales 
between the upper two panels $\mu=0.2,0.4$ and the lower two
$\mu=0.6,0.8$. For each $\mu$, ${\cal C}_{\gamma_5}(\vec k=\vec0)$ 
is almost flat, indicating a massless
state.

Appendix B sketches the calculation of ${\cal C}_\Gamma(\vec k,x_0)$ as a 
function of $\mu$ in free field theory. As $x_0\to\infty$ the function
is dominated by a continuum of particle-hole pairs 
at or near the Fermi surface, which
effectively cost zero energy to excite. The generic result is that for
$\vert\vec k\vert\leq2\mu$ the decay is algebraic, with
(\ref{eq:c5},\ref{eq:Cbigk})
\begin{equation}
{\cal C}_\Gamma(\vec k,x_0)\propto\cases{x_0^{-2},& $\vert\vec k\vert\ll\mu$;\cr
     x_0^{-{3\over2}},&$\vert\vec k\vert\simeq2\mu$.\cr}
\end{equation}
Only once $\vert\vec k\vert>2\mu$ does it become kinematically impossible to
excite a pair with zero energy, resulting in exponential decay:
\begin{equation}
{\cal C}_\Gamma(\vec k,x_0)\propto x_0^{-{3\over2}}\exp\Bigl(-(\vert\vec
k\vert-2\mu)x_0\Bigr).
\end{equation}
The sequence of plots in Fig.~\ref{fig:psdisperse} is in qualitative agreement
with these findings. For each $\mu$ there is a particular value of
$\vert\vec k\vert$, highlighted with a solid line in the plots, for which the 
temporal falloff is particularly slow, corresponding to 
$\vert\vec k\vert\approx2\mu$ (eg. for $\mu=0.4$ the slow falloff 
occurs for $\vert\vec k\vert={\pi\over4}\simeq0.785$). 
For $\vert\vec k\vert$
larger than this value the decay
is much steeper, although only for $\mu=0.2$ does it resemble an
exponential form.
Because of the technical difficulties in treating correlators with power-law 
decays in a finite volume (see eg. \cite{HKK1}) we have made no attempt to
fit the numerical data for ${\cal C}_{\gamma_5}(\vert\vec k\vert,x_0)$ 
to a functional form.
It can be observed, however, that the overall magnitude of ${\cal C}_{\gamma_5}$
increases with $\mu$ as the size of the Fermi surface and hence the number of
participating particle-hole states grows. There is, however,
no obvious correspondence
with the prefactor $\vert\vec k\vert/\mu$ predicted in (\ref{eq:c5}), probably
because the approximation $\vert\vec k\vert\ll\mu$ is not valid over the
accessible momentum range. 

\begin{figure}[]
\bigskip\bigskip
\begin{center}
\epsfig{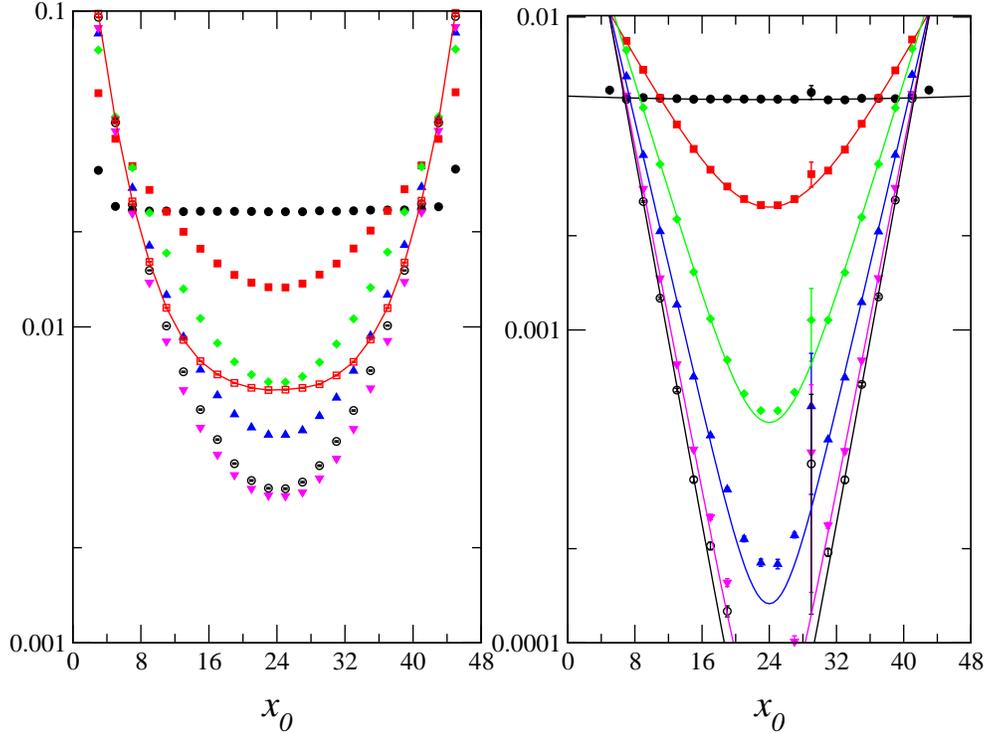}
\end{center}
\caption{Vector correlators ${\cal C}_{\gamma_\perp}(\vert\vec k\vert,x_0)$ 
(left) and
${\cal C}_{\gamma_\parallel}(\pi-\vert\vec k\vert,x_0)$ (right) at $\mu=0.6$. 
The symbols have the same
meaning as in Fig.~\ref{fig:psdisperse}.}
\label{fig:rhodisp}
\end{figure}

\begin{figure}[]
\bigskip\bigskip
\begin{center}
\epsfig{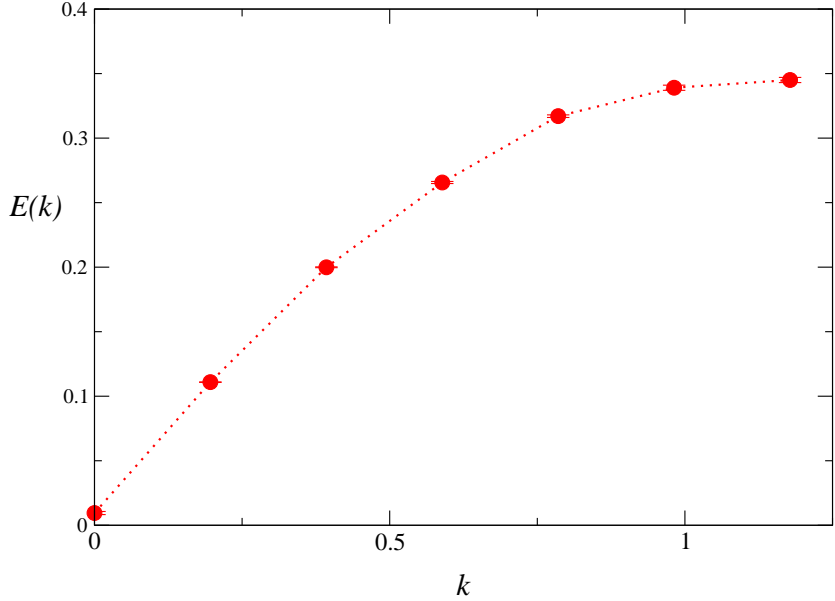}
\end{center}
\caption{Dispersion relation $E(\vert\vec k\vert)$ extracted from 
${\cal C}_{\gamma_\parallel}$
.}
\label{fig:zerosound}
\end{figure}

The situation is more interesting in the vector channel, corresponding to
the quantum numbers of the $\rho$ meson, because for $\vert\vec k\vert>0$ 
we can distinguish
between ${\cal C}_{\gamma_\parallel}(\vec k)$, 
in which the component of the vector is parallel
to $\vec k$, and ${\cal C}_{\gamma_\perp}(\vec k)$. 
In Fig.~\ref{fig:rhodisp} we plot 
the correlator for several $\vert\vec k\vert$ values at $\mu=0.6$ in each case. 
For ${\cal C}_{\gamma_\perp}$, just as for ${\cal C}_{\gamma_5}$,
there is no difference between the data for $\vert\vec k\vert$ and
$\pi-\vert\vec k\vert$. For
${\cal C}_{\gamma_\parallel}$, however, because of the point-split nature of the
operator $j_{\gamma_i}$ this symmetry no longer holds; in this case we
choose to plot data with $\vert\vec k\vert\in[{\pi\over2},\pi]$ (the data for
$\vert\vec k\vert\leq{\pi\over2}$ are very similar to those for
${\cal C}_{\gamma_\perp}$). 
Each plot covers two decades on the vertical axis; however
both the magnitude and the shape of the curves is very different.
For ${\cal C}_{\gamma_\perp}$ 
the curves are qualitatively very similar to those of 
Fig.~\ref{fig:psdisperse} at $\mu=0.6$, with a distinguished momentum 
$\vert\vec k\vert={3\pi\over8}$. 
The correlator ${\cal C}_{\gamma_\parallel}$ is much smaller in
magnitude, and is consistent with exponential rather than algebraic decay. The
lines are fits of the form ${\cal C}_{\gamma_\parallel}(\vert\vec k\vert,x_0)=
A(e^{-Ex_0}+e^{-E(L_t-x_0)})$. The resulting $E(\vert\vec k\vert)$ is shown in
Fig.~\ref{fig:zerosound}. For small $\vert\vec k\vert$ 
it resembles that of a massless
pole, ie. $E=\beta_0\vert\vec k\vert$, with velocity $\beta_0\approx0.5$.
It should be mentioned, however, that although ${\cal C}_{\gamma_\parallel}$ and
${\cal C}_{\gamma_\perp}$ still differed, no evidence for a massless pole
was seen in the data at $\mu=0.8$.

Light states in the $\rho$ channel in medium have long been of interest
\cite{BRHL}. One recently proposed theoretical
scenario is that the $\rho$ should be
viewed as an ``almost'' Goldstone boson due to a spontaneous ``induced''
breaking of Lorentz symmetry over and above the explicit breaking due to
$\mu\not=0$ \cite{LRR}. In this case the $\rho$ mass is predicted to scale
as $m_\rho^2\propto2\mu n$. Another, proposed in the context of Two Color QCD
(TCQCD) in
which mesons and baryons fall in the same multiplets and are degenerate at
$\mu=0$, is that for sufficiently large $\mu$ there is a condensation of 
vector diquark states leading to the spontaneous 
breaking of rotational invariance and 
vector Goldstone states with both $E\propto\vert\vec k\vert$ 
and $E\propto \vert\vec k\vert^2$ \cite{SS}.
In-medium modification of the $\rho$ correlator has recently been observed in 
lattice Monte Carlo simulations of TCQCD \cite{MNN}.

Since our results suggest a massless state 
and there is no Fermi surface in
TCQCD we favour an alternative explanation, namely that the pole signals
a collective excitation of the system in which the shape of the Fermi surface
is distorted, leading to a rotationally non-invariant 
disturbance propagating at velocity $\beta_0$
known as zero sound. The conditions for zero sound to propagate 
are that there must be interactions, ie. 
${\cal F}_{\vec k,\vec k^{\prime}}\not=0$,
and that inter-particle collisons are negligible, ensured if $T\approx0$
\cite{Landau3,Landau2}. Indeed, zero sound solutions can in principle 
be found self-consistently using the
theoretical framework of Sec.~\ref{sec:fermiliquid}, although the specific form
(\ref{eq:interaction}) for the Fermi liquid interaction entails the solution
of an integral equation of the second kind, beyond the scope of this
paper. A potential problem for this picture is that the extracted velocity 
$\beta_0\approx0.5$ is less than
$\beta_F\approx1$, implying that quasiparticles would experience damping via 
\v Cerenkov radiation of zero sound. Clearly further work exploring the
systematic
effects of varying $\mu$, $\Sigma_0$, $\vec k$ and volume  
will be needed for a more complete understanding to emerge.

\subsection{Mesonic Wavefunctions and Friedel Oscillations}
\label{sec:friedel_num}

\begin{figure}[]
\bigskip\bigskip
\begin{center}
\epsfig{file=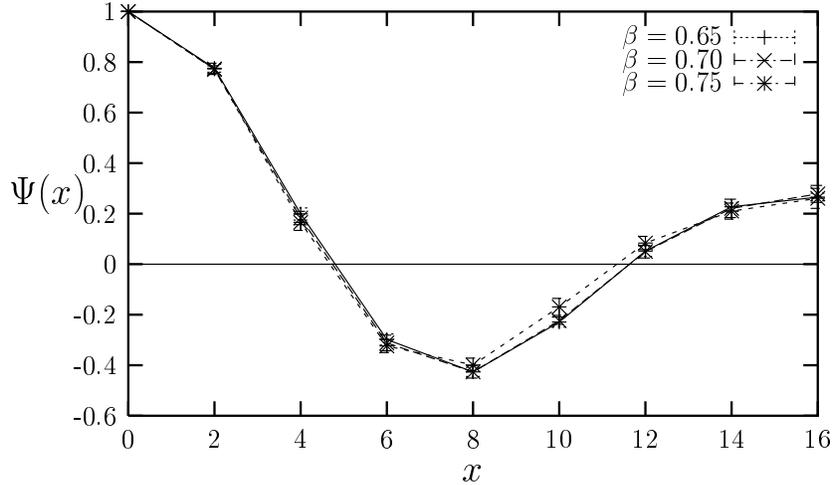, width=11cm}
\end{center}
\caption{Scalar wavefunction at $\mu=0.5$ for $\beta=0.65, 0.70, 0.75$.}
\label{fig:discret}
\end{figure}

We now switch attention to spatial correlations between particles,
probed via the wavefunction $\Psi(\vec x)$ defined by
\begin{equation}
\Psi_\Gamma(\vec x)=\lim_{x_0\to\infty}\Psi_\Gamma^{-1}(\vec x=\vec0)
\sum_{\vec y}\langle{\cal G}_q(\vec 0,0;\vec
y,x_0)\Gamma{\cal G}_{\bar q}(\vec 0,0;\vec y+\vec x,x_0)\Gamma\rangle
\end{equation}
where as ususal 
$\Gamma$ projects out the quantum numbers of the channel of interest.
In the GN and related models $\Psi(\vec x)$ is technically
much easier to define and measure than in QCD-like theories where they are 
gauge-dependent.
Meson wavefunctions in the 2+1$d$ GN model have been studied at $T,\mu=0$ in 
\cite{HKS}, where further technical details are given.

\begin{figure}[]
\bigskip\bigskip
\begin{center}
\epsfig{file=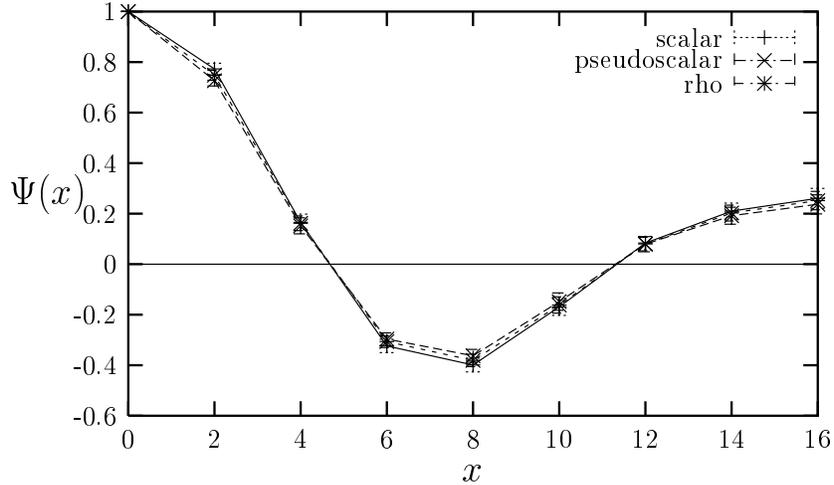, width=11cm}
\end{center}
\caption{$\Psi(\vert\vec x\vert)$ 
for scalar, pseudoscalar and vector channels
at $1/g^2=0.75$, $\mu=0.5$.}
\label{fig:all.mesons}
\end{figure}

Fig.~\ref{fig:discret} shows $\Psi(\vec x)$ measured along the $x$-axis in the 
scalar channel at a constant $\mu=0.5$ at various 
values of the the coupling $1/g^2=0.65, 0.70, 0.75$. The oscillatory behaviour
described in Sec.~\ref{sec:friedel} is clearly seen, and is plainly  
not a discretisation artifact since its form is stable as $a(g)$ varies.
Fig.~\ref{fig:all.mesons} shows that there are no significant differences among
the various
mesonic channels implying that in contrast to the situation at $\mu=0$
\cite{HKS}, effects due to eg. $\sigma$ exchange are very hard to detect.

In the next four figures we plot the scalar wavefunctions at $1/g^2=0.75$ for 
a sequence of $\mu$ values 0.2,
0.4, 0.6, and 0.8.
The simulation data are connected by solid lines. As $\mu$, and hence
$k_F$, increases the
oscillations decrease in wavelength in accordance with theoretical
expectation, and provide a graphic illustration of the presence of a sharp Fermi
surface. The dashed lines show measurements taken with the same
lattice parameters but with the
interaction switched off. The disparity with the interacting theory is 
small, though increasing with $\vert\vec x\vert$, 
showing that the free field description of
the oscillations is qualitatively correct. A possible explanation 
is the infinite value predicted for the
Debye mass in Sec.~\ref{sec:aux}, implying that interactions between static 
quarks at non-zero $\vert\vec x\vert$ are completely screened. 
Also shown is the 
theoretical form (\ref{eq:friedel})
$\Psi(x)=J_0(k_F x)$ with $k_F\equiv\mu$, showing good agreement with the data 
for small $x\lapprox2k_F^{-1}$. Unfortunately it appears hard to
obtain more quantitative information, such as an independent fit for
the Fermi momentum $k_F$, because $\vert J_0(kx)\vert$
decays only as $x^{-{1\over2}}$. This means that fits should
not only include the backwards-propagating signal $J_0(k_F(L_s-x))$ but also 
image contributions $J_0(k_F(nL_s-x))$ \cite{HKK1} -- our attempts to find a
satisfactory fit were unsuccessful. The figures therefore 
simply show both the ``forwards'' $J_0(\mu x)$
and ``forwards-and-backwards'' $J_0(\mu x) + J_0(\mu(L_s-x))$ forms, showing 
that neither gives a satisfactory description of the data over the full range 
of $x$ and $\mu$. It is also possible that more sophisticated fits taking proper
account of discretisation effects are needed; at $\mu=0.8$ the baryon density
$n=0.31$ (see Table~\ref{tab:n}), a significant fraction of its saturation value
of 1, so that the Fermi surface may be distorted.

\begin{figure}[]
\bigskip\bigskip
\begin{center}
\epsfig{file=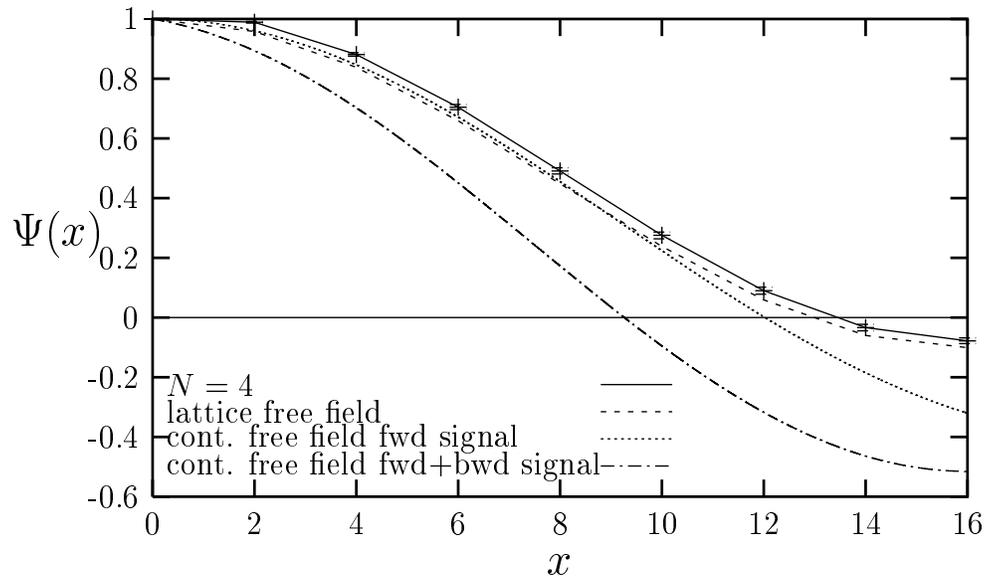, width=13cm}
\end{center}
\caption{Scalar wavefunction at $\mu=0.2$.}
\label{fig:wvfn2}
\end{figure}

\begin{figure}[]
\bigskip\bigskip
\begin{center}
\epsfig{file=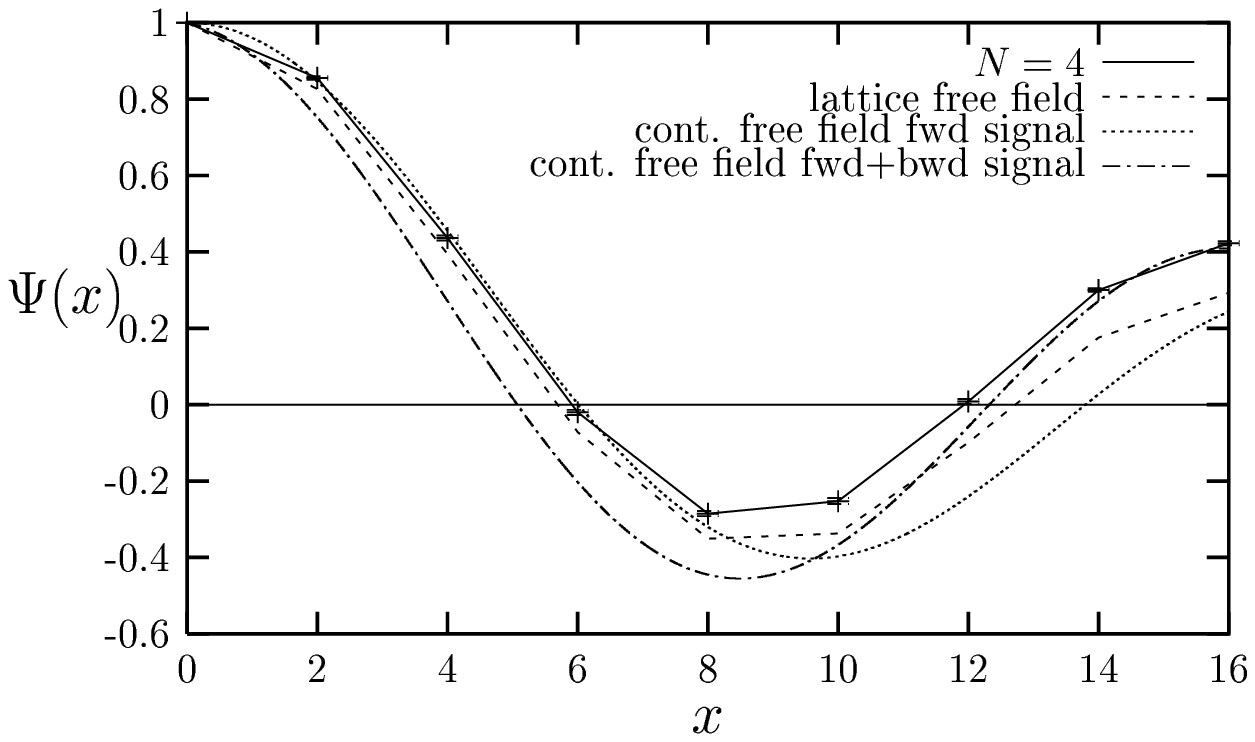, width=13cm}
\end{center}
\caption{Scalar wavefunction at $\mu=0.4$.}
\label{fig:wvfn4}
\end{figure}

\begin{figure}[]
\bigskip\bigskip
\begin{center}
\epsfig{file=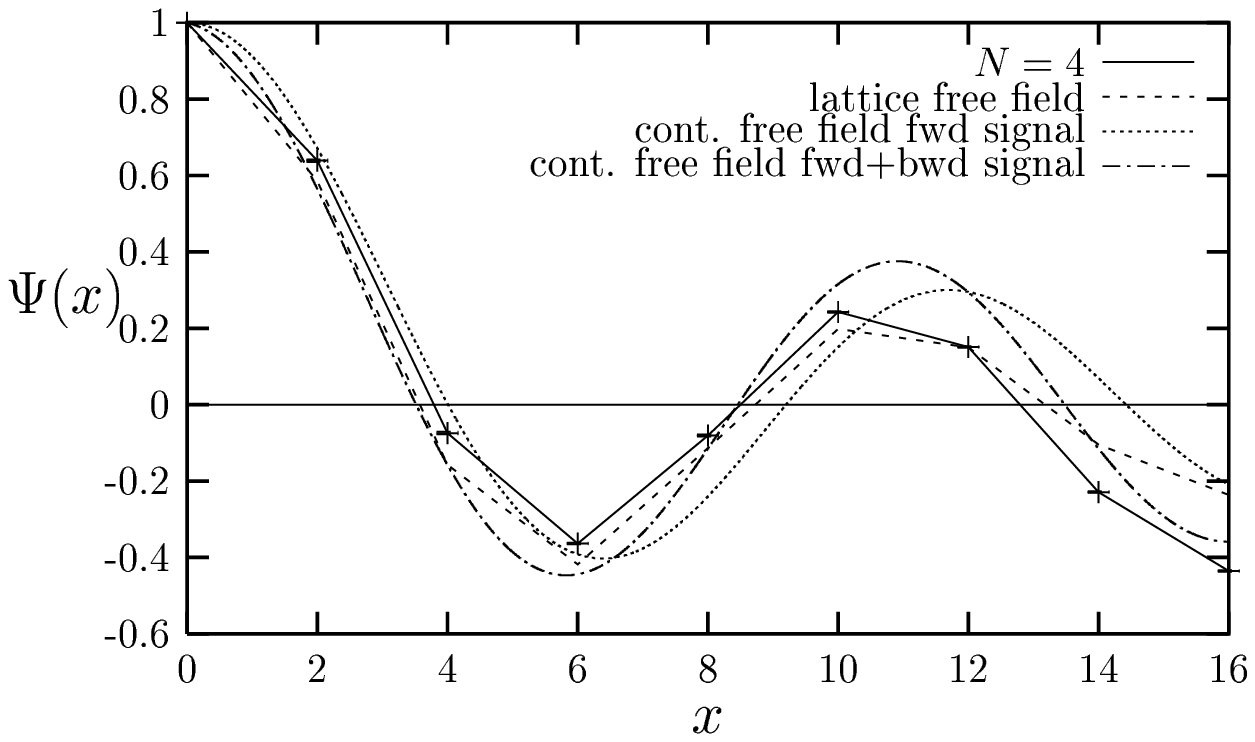, width=13cm}
\end{center}
\caption{Scalar wavefunction at $\mu=0.6$.}
\label{fig:wvfn6}
\end{figure}

\begin{figure}[]
\bigskip\bigskip
\begin{center}
\epsfig{file=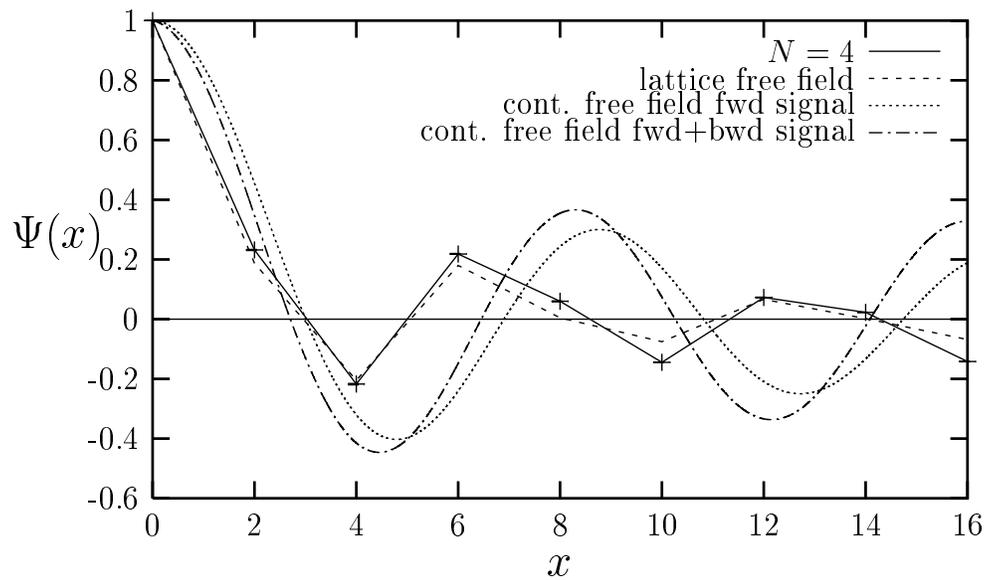, width=13cm}
\end{center}
\caption{Scalar wavefunction at $\mu=0.8$.}
\label{fig:wvfn8}
\end{figure}

\section{Summary}

The GN and related models remain the only interacting 
field theories both simulable by
standard lattice methods at $\mu\not=0$ and displaying a Fermi surface, thought
to be an essential feature of dense quark matter. In this paper we have
attempted to develop an understanding of how orthodox lattice observables are 
altered in such conditions, and how they should be interpreted using the
language of many-body physics.
The main achievements have been:

\begin{itemize}

\item
An analytic calculation of the auxiliary boson propagator $D_\sigma$ in medium
to leading order in $N_f^{-1}$ in the HDL approximation.
The branch cut in the complex-$k$ plane is modified to become an isolated pole.
Physically, the Debye mass is infinite, but the plasma frequency finite and
vanishing as onset is approached. The dispersion relation $E(k)$ is non-trivial.
Lattice simulation has verified these results with acceptable precision (ie. to
within $O(N_f^{-1}$)), and have unambiguously identified in-medium effects in
this channel.

\item
A systematic calculation of Fermi liquid parameters up to $O(N_f^{-1})$, which
is consistent with causality constraints and  has
in turn been verified by lattice simulation. This contrasts with existing
studies of the quasiparticle dispersion relation in the 2+1$d$ NJL model with
$\mu>\mu_c$, which
are {\em not\/} consistent with the large-$N_f$ predictions.

\item
A study of meson correlation functions at $\vec k\not=0$; instead of showing 
exponential falloff with Euclidean time these generically decay algebraically, 
signalling the presence of massless particle-hole 
excitations for all $\vec k$. There is 
qualitative agreement between the results of free field theory and lattice
simulation, which both show a non-trivial dependence on $\vert\vec k\vert/\mu$.
There is also tentative evidence for a massless pole in the vector channel,
which is possibly a manifestation of zero sound.

\item
The first observation of oscillatory behaviour in mesonic wavefunctions, which
resemble the Friedel oscillations familiar in many-body physics.
Figs.~\ref{fig:wvfn2}-\ref{fig:wvfn8} offer a graphic confirmation, if one is
still needed, of the presence of a sharp Fermi surface in this model.

\end{itemize}

It would be interesting to apply the same analysis to the lattice
NJL model at $\mu\not=0$, which has
been studied in both 2+1$d$ \cite{HLM} and 3+1$d$ \cite{HW}. Independent of
the 
issue of whether the ground state has a non-vanishing diquark condensate, these
models are more realistic because they contain physical pions, whose behaviour
at the quark-hadron transition is thought to change dramatically. Indeed,
the result of Sec.~\ref{sec:fermiliquid} that the plasma frequency (pole mass)
in the chirally symmetric phase
decreases as $\mu\to\mu_{c+}$ while the Debye (screening) mass diverges is 
reminiscent of recent results obtained by Son and Stephanov in the chirally
broken phase \cite{SoSt} using an effective theory and scaling arguments.
It would be straightforward to repeat our analysis for the pion dispersion
relation. More generally, it is worth noting that should an algorithm effective
at $\mu\not=0$ ever emerge our results in the meson sector
are in principle accessible as gauge invariant correlators
in a QCD simulation. The $\sigma$ channel in the current model 
is dominated by disconnected $q\bar q$ bubbles which are technically hard 
to compute if an auxiliary field is absent, but in the lightest channel in QCD,
namely the pion,  the relevant diagrams have connected quark lines. The main
difference in the baryon sector, of course, is that gauge invariant
quasiparticles now
have $qqq$ quantum numbers, and at least below the transition should be 
non-relativistic at the Fermi surface.

\section*{Acknowledgements}

SJH was supported by a PPARC Senior Research Fellowship, CGS by the
Leverhulme Trust and DOE grant DE-FG02-96ER40945, and JBK by NSF grant
NSF-PHY-0102409. 
We are grateful to Kurt Langfeld for discussions in the early
stages of this work.

\section*{Appendix A}

In this appendix we outline how oscillatory behaviour as a function of spatial
separation $\vec y$ resembling Friedel
oscillations appears in the large Euclidean time $x_0\to\infty$ limit
of the density-density correlator 
$C_{nn}(\vec y;x_0)=\sum_{\vec x}\langle j_0(\vec0,0)j_0^\dagger(\vec x,x_0)
j_0(\vec0,0)j_0^\dagger(\vec x+\vec y,x_0)\rangle$, where 
$j_0(x)=\bar\psi\gamma_0\psi(x)$. The relevant expression is
\begin{equation}
\sum_{\vec x}\mbox{tr}\int_p\int_q\int_k\int_\ell\left\{
\gamma_0{{e^{ipx}}\over{ip{\!\!\! /\,}+\mu\gamma_0}}
\gamma_0{{e^{-iqx}}\over{iq{\!\!\! /\,}+\mu\gamma_0}}
\gamma_0{{e^{ikx}e^{i\vec k.\vec y}}\over{ik{\!\!\! /\,}+\mu\gamma_0}}
\gamma_0{{e^{-i\ell x}e^{-i\vec\ell.\vec y}}\over{i\ell{\!\!\! /\,}
+\mu\gamma_0}}\right\},
\end{equation}
where we have assumed
massless quark propagation in the high density phase. The trace yields three
terms of which the first is
\begin{equation}
\int_p\int_q {{[\vec p.\vec q-(p_0-i\mu)(q_0-i\mu)]e^{i(p-q)x}}\over
{[(p_0-i\mu)^2+\vec p^2][(q_0-i\mu)^2+\vec q^2]}} \times
\int_k\int_\ell {{[\vec k.\vec \ell-(k_0-i\mu)(\ell_0-i\mu)]e^{i(k-\ell)x}
e^{i(\vec k-\vec\ell).\vec y}}\over
{[(k_0-i\mu)^2+\vec k^2][(\ell_0-i\mu)^2+\vec \ell^2]}} 
\end{equation}
The integrals over timelike momenta are performed by completing the 
contour in the upper half-plane for $\int dp_0$, $\int dk_0$ and in the lower
half-plane for $\int dq_0$, $\int d\ell_0$, noting the pole positions at
eg. $p_0=i\mu\pm i\vert\vec p\vert$. In fact, only the pole with negative
imaginary part contributes to $\int dq_0$ and $\int d\ell_0$, and it is
straightforward to show that the dominant term for large $x_0$ in the resulting
expression 
comes from the poles at $p_0=i\mu-i\vert\vec
p\vert$ and $k_0=i\mu-i\vert\vec k\vert$
since all others are suppressed by powers of $e^{-\mu x_0}$. 
The resulting contribution
from all three 
trace terms is 
\begin{eqnarray}
\sum_{\vec x}{1\over4}
\int_0^\mu{{d^dp}\over{(2\pi)^d}}
\int_\mu^\infty{{d^dq}\over{(2\pi)^d}}
\int_0^\mu{{d^dk}\over{(2\pi)^d}}
\int_\mu^\infty{{d^d\ell}\over{(2\pi)^d}}
e^{i(\vec p-\vec q+\vec k-\vec\ell).\vec x}
e^{-(\vert\vec q\vert-\vert\vec p\vert+
\vert\vec\ell\vert-\vert\vec k\vert)x_0}e^{i(\vec k-\vec\ell).\vec y}\times
\nonumber\\
\left\{(1+\cos\theta_{pq})(1+\cos\theta_{k\ell})-
(1-\cos\theta_{pk})(1-\cos\theta_{q\ell})+
(1+\cos\theta_{p\ell})(1+\cos\theta_{qk})\right\},\label{eq:bigint}
\end{eqnarray}
where $\theta_{pq}$ denotes the angle between vectors $\vec p$ and $\vec q$,
etc.

Now, the sum over the timeslice $\sum_{\vec x}$ results in an overall
momentum-conserving delta-function.
Since, however, the integrand is dominated by momenta in the immediate 
vicinity of the 
Fermi surface, a physically reasonable approximation, which 
preserves all angles under the remaining $\int d^dp\int d^dq$, 
is given by 
\begin{equation}
(2\pi)^d\delta^d(\vec p-\vec q+\vec k-\vec\ell)\sim
(2\pi)^d\delta^d(\vec p+\vec k)(2\pi)^d\delta^d(\vec q+\vec\ell).
\end{equation}
The expression (\ref{eq:bigint}) reduces to
\begin{equation}
\int_0^\mu{{d^dp}\over{(2\pi)^d}}
\int_\mu^\infty{{d^dq}\over{(2\pi)^d}}e^{-2(\vert\vec q\vert-\vert\vec
p\vert)x_0}e^{-i(\vec p-\vec q).\vec y}
{1\over2}(\cos^2\theta_{pq}-1).
\end{equation}
Denote the angle between $\vec p$ ($\vec q$) and $\vec y$ by $\theta_1$
($\theta_2$). Then
\begin{equation}
\cos\theta_{pq}=\cos\theta_1\cos\theta_2
+\sin\theta_1\sin\theta_2\cos\Theta_{pq},
\end{equation}
where $\Theta_{pq}$ is the angle between the vectors projected onto the
subspace $\Omega^{d-1}$ orthogonal to $\vec y$. 
The angular integrals over this subspace may be performed using
\begin{equation}
\int d\Omega^{d-1}={{2\pi^{(d-1)/2}}\over{\Gamma({{d-1}\over2})}}\;\;\;;
\;\;\;\int d\Omega^{d-1}\cos^2\Theta={1\over{d-1}}
{{2\pi^{(d-1)/2}}\over{\Gamma({{d-1}\over2})}},\nonumber
\end{equation}
to yield
\begin{eqnarray}
{1\over{(2\pi)^{2d}}}{{2\pi^{d-1}}\over{\Gamma^2({{d-1}\over2})}}
\int_0^\mu p^{d-1}dp\int_\mu^\infty q^{d-1}dq \;e^{-2(q-p)x_0}
\int_0^\pi d\theta_1\int_0^\pi d\theta_2 
e^{-ipy\cos\theta_1}e^{iqy\cos\theta_2}\times\nonumber\\
\sin^{d-2}\theta_1\sin^{d-2}\theta_2
\left\{\cos^2\theta_1\cos^2\theta_2+{1\over{d-1}}\sin^2\theta_1\sin^2\theta_2
-1\right\}.
\end{eqnarray}

The remaining angular integrals are readily performed to yield
\begin{eqnarray}
&\displaystyle{{(d-1)}\over{2(2\pi)^d}}&
\!\!\!\!\int_0^\mu p^{d-1}dp\int_\mu^\infty q^{d-1}dq
\;e^{-2(q-p)x_0}\times\nonumber\\
&\displaystyle{\left({1\over{\sqrt{pq}y}}\right)^{d-2}}&\!\!\!\!\!\!\left\{
-{1\over{py}}J_{d\over2}(py)J_{{d-2}\over2}(qy)-
       {1\over{qy}}J_{d\over2}(qy)J_{{d-2}\over2}(py)
+{d\over{pqy^2}}J_{d\over2}(py)J_{d\over2}(qy)\right\}.
\end{eqnarray}
The integrals over vector lengths
$p$ and $q$ are estimated using Laplace's method;
\begin{equation}
\lim_{x_0\to\infty}\int_0^\mu dp p^{d-1}pe^{2px_0}f(py)
\int_\mu^\infty dq q^{d-1}
e^{-2qx_0}g(qy)\sim{\mu^{2d-2}\over{4x_0^2}}f(\mu y)g(\mu
y)\left(1+O(x_0^{-1})\right),
\end{equation}
to yield the final result
\begin{equation}
{{(d-1)}\over{8(2\pi)^d}}{{\mu^{2d-2}}\over{x_0^2(\mu y)^{d-2}}}
\left[-{2\over{\mu y}}J_{{d-2}\over2}(\mu y)J_{d\over2}(\mu y)+
       {d\over{(\mu y)^2}}J_{d\over2}^2(\mu y)\right].
\end{equation}
One may now use the asymptotic form of the Bessel function to deduce the 
long-distance behaviour of $C_{nn}(\vert\vec y\vert;x_0)$:
\begin{equation}
C_{nn}(y;x_0)\sim\cases{{1\over{16\pi^3 x_0^2}}{{\cos2\mu y}\over{y^2}}, &
$d=2$;\cr
{{\mu}\over{16\pi^4x_0^2}}{{\sin2\mu y}\over{y^3}}, &$d=3$. \cr}
\end{equation}

\section*{Appendix B}

In this section we analyse the behaviour of mesonic correlators 
as a function of spatial momentum Euclidean time in the
high density phase. For simplicity we fix $d=2$.
The general expression is
\begin{equation}
{\cal C}_\Gamma(\vec k;x_0)=\sum_{\vec x}\int_p\int_q
{{\mbox{tr}\{\Gamma(-i\vec p.\vec\gamma-i(p_0-i\mu)\gamma_0)\Gamma
(-i\vec q.\vec\gamma-i(q_0-i\mu)\gamma_0)\}}\over
{[(p_0-i\mu)^2+\vert\vec p\vert^2][(q_0-i\mu)^2+\vert\vec q\vert^2]}}
e^{i(p-q)x}e^{i\vec k.\vec x},
\end{equation}
where we will consider $\Gamma\in\{\One,\gamma_5,\gamma_\parallel,\gamma_\perp,
\gamma_o\}$, $\parallel$ and $\perp$ denoting components parallel and
perpendicular to $\vec k$. On
performing the trace, the timeslice sum $\sum_{\vec x}$ and the timelike
momentum integrals, we are left with, eg.
\begin{equation}
{\cal C}_{\gamma_5}(\vec k;x_0)=\int{{d^2\vec p}\over{(2\pi)^2}}
\theta(\mu-\vert\vec p\vert)\theta(\vert\vec p+\vec k\vert-\mu)
{{\vert\vec p\vert\vert\vec p+\vec k\vert-\vec p.(\vec p+\vec k)}\over
{\vert\vec p\vert\vert\vec p+\vec k\vert}}
e^{-(\vert\vec p+\vec k\vert-\vert\vec p\vert)x_0}
\label{eq:corr}
\end{equation}
where we have assumed $x_0>0$ and once again focussed on the contribution
from the dominant pole $p_0=i(\mu-\vert\vec p\vert)$. We can evaluate the
remaining integral in two limits:

\medskip
\noindent{\Large{$\vert\vec k\vert\ll\mu\,$:}}

In this case the integration region is a narrow
crescent in the vicinity of the Fermi surface, and the integrand and limits
can be expanded
in powers of ${\vert\vec k\vert/\vert\vec p\vert}$. The general form is then
\begin{equation}
{\cal C}_\Gamma(\vec k;x_0)
=2\mu\vert\vec k\vert\int_{-{\pi\over2}}^{\pi\over2}{{d\theta}\over{2\pi}}
h_\Gamma(\theta)\exp(-\vert\vec k\vert x_0\cos\theta)
\biggl(1+O\Bigl({{\vert\vec k\vert}\over\mu}\Bigr)\biggr)
\end{equation}
where the leading contribution for each channel is given by
\begin{eqnarray}
h_{\gamma_5}=-h_{\One}={{\vert\vec
k\vert^2}\over{4\mu^2}}\cos\theta\sin^2\theta;\;\;\;\;\;
h_{\gamma_\perp}=\cos\theta\sin^2\theta;
\;\;\;\;
h_{\gamma_0}=-\cos\theta;
\nonumber\\
h_{\gamma_\parallel}=\cos^3\theta+{{3\vert\vec
k\vert}\over{2\mu}}\cos^2\theta\sin^2\theta+
{{\vert\vec k\vert^2}\over{4\mu^2}}\cos\theta\sin^2\theta(1+2\sin^2\theta).
\end{eqnarray}
The angular integral leads to the following results for the leading behaviour
in $\vert\vec k\vert/\mu$:
\begin{eqnarray}
{\cal C}_{\gamma_5,\One}(\vec k;x_0)&=&\!\!\!
\pm{{\vert\vec k\vert^3}\over{2\pi\mu}}\biggl[{1\over3}+{\pi\over2}
\biggl({{{\bf L}_2(\vert\vec k\vert x_0)-I_{-2}(\vert\vec k\vert x_0)}\over
{\vert\vec k\vert x_0}}
\biggr)\biggr] \sim \pm{{\vert\vec k\vert}\over{2\pi\mu x_0^2}}+O(x_0^{-4})
\label{eq:c5}\\
{\cal C}_{\gamma_\perp}(\vec k;x_0)
&=&{{4\mu^2}\over{\vert\vec k\vert^2}}{\cal C}_{\gamma_5}(\vec k;x_0)
\sim{{2\mu}\over{\pi\vert\vec k\vert x_0^2}}+O(x_0^{-4})\\
{\cal C}_{\gamma_\parallel}(\vec k;x_0)
&=&{{11\vert\vec k\vert^3}\over{30\pi\mu}}
+{{3\vert\vec k\vert^3}\over{2\mu}}\biggl(
{{{\bf L}_3(\vert\vec k\vert x_0)-I_{-3}(\vert\vec k\vert x_0)}\over
{(\vert\vec k\vert x_0)^2}}
\biggr)\\
+\vert\vec k\vert^2\biggl(
{{\vert\vec k\vert}\over{4\mu}}\!\!\!\!\!&-&\!\!\!\!\!
{3\over2}{{d\;}\over{d\beta}}\biggr)\biggl(
{{{\bf L}_2(\beta)-I_{-2}(\beta)}\over\beta}
\biggr)_{\beta=\vert\vec k\vert x_0}
\!\!\!\!+\mu\vert\vec k\vert{{d^2\;}\over{d\beta^2}}\biggl(
{\bf L}_1(\beta)-I_{-1}(\beta)\biggr)_{\beta=\vert\vec k\vert x_0}\nonumber\\
&\sim&
{{3\vert\vec k\vert}\over{2\pi\mu x_0^2}}+{6\over{\pi\vert\vec k\vert x_0^3}}+
{{12\mu}\over{\pi\vert\vec k\vert^3 x_0^4}}
+O\Bigl((\mu\vert\vec k\vert x_0^4)^{-1},(\vert\vec k\vert^3x_0^5)^{-1},
\mu(\vert\vec k\vert^5x_0^6)^{-1}\Bigr)\nonumber\\
{\cal C}_{\gamma_0}(\vec k;x_0)&=&\!\!\!-{{2\mu\vert\vec k\vert}\over\pi}
\biggl[1+{\pi\over2}\Bigl({\bf L}_1(\vert\vec k\vert x_0)-I_{-1}(\vert\vec
k\vert x_0)\Bigr)\biggr]
\sim-{{2\mu}\over{\pi\vert\vec k\vert x_0^2}}+O(x_0^{-4})
\label{eq:c0}
\end{eqnarray}
where $I$ and {\bf L} are respectively modified Bessel and Struve
functions. Fortunately their difference is a tabulated function \cite{AbSteg},
which enables the asymptotic form of the second equality in eqns. (\ref{eq:c5} -
\ref{eq:c0}) to be determined. The decay in the timelike direction
is generically $\propto x_0^{-2}$, 
the interesting exception being for $\Gamma=\gamma_\parallel$ 
where there are significant corrections at shorter distances.

\medskip
\noindent{\Large$\vert\vec k\vert\ge2\mu\,$:}

In this case the $\theta$ functions
in (\ref{eq:corr}) yield unity for all $\vert\vec p\vert\leq\mu$, simplifying
the integration limits. We will restrict analysis to the case $\Gamma=\gamma_5$:
\begin{eqnarray}
&&\int_0^{2\pi}{{d\theta}\over{2\pi}}\int_0^\mu pdp\left[
1-{{p^2+p\vert\vec k\vert\cos\theta}\over{p^2(1+{{2\vert\vec k\vert}\over
p}\cos\theta+{{\vert\vec k\vert^2}\over p^2})^{1\over2}}}\right]
e^{-{\displaystyle px_0}\Bigl[\sqrt{\textstyle{1+{{2\vert\vec k\vert}\over
p}\cos\theta+{{\vert\vec k\vert^2}\over p^2}}}-{\displaystyle1}\Bigr]}
=\nonumber\\
&&{\mu^2\over\pi}\int_{-\infty}^\infty du\int_0^1dv
v\left[1-
{{1+{{v+\kappa}\over{v-\kappa}}u^2}\over
{[(1+u^2)(1+({{v+\kappa}\over{v-\kappa}})^2u^2)]^{1\over2}}}\right]
{{e^{-{\displaystyle\mu x_0}
\Bigl[\sqrt{\textstyle{{(v-\kappa)^2+(v+\kappa)^2u^2}\over{1+u^2}}}
-{\displaystyle v}
\Bigr]}}\over{1+u^2}},\nonumber\\
\end{eqnarray}
where in the second line we have used the substitutions
$u=\cot{1\over2}\theta$, $v=p/\mu$, and $\kappa=\vert\vec k\vert/\mu\geq2$.
To focus on the large-$x_0$ behaviour 
we now employ Laplace's method, expanding the integrand in powers of $u^2$
and noting that since $v-\kappa<0$ the negative branch of the square roots
should be chosen.
The dominant contribution is 
\begin{equation}
{{2\mu^2}\over\pi}\int_0^1dvve^{-\mu x_0(\kappa-2v)}
\int_{-\infty}^\infty du
e^{-\mu x_0{{2\kappa v}\over{\kappa-v}}u^2}
={\mu^2\over{\surd\pi}}\left({2\over{\kappa\mu x_0}}\right)^{1\over2}
e^{-\kappa\mu x_0}\int_0^1dvv^{1\over2}(\kappa-v)^{1\over2}e^{2\mu x_0v}.
\end{equation}
The final integral is estimated by replacing the factor
$(\kappa-v)^{1\over2}$, which is non-singular within the integration range,
by $(\kappa-1)^{1\over2}$;
we obtain the lower bound
\begin{equation}
{\cal C}_{\gamma_5}(\kappa;x_0)\geq{\mu^2\over{\surd\pi}}
\left({{2(\kappa-1)}\over{\kappa\mu x_0}}\right)^{1\over2}
e^{-\kappa\mu x_0}(-2\mu x_0)^{-{3\over2}}\gamma(\textstyle{3\over2},-2\mu x_0)
\end{equation}
where $\gamma$ is the incomplete Gamma function, whose asymptotic
form for large $\mu x_0$ yields the final answer
\begin{equation}
{\cal C}_{\gamma_5}(\vec k;x_0)\geq{\mu^2\over{2\surd\pi}}
{{e^{-(\vert\vec k\vert-2\mu)x_0}}\over{(\mu x_0)^{3\over2}}}
\biggl(1+O(\mu x_0)^{-1}\biggr).
\label{eq:Cbigk}
\end{equation}
For $\vert\vec k\vert>2\mu$ the correlator thus decays exponentially; however
for the critical value $\vert\vec k\vert=2\mu$ the decay is algebraic, and
moreover slower than that for $\vert\vec k\vert\ll\mu$
(\ref{eq:c5}).


\begin{thebibliography}{xx}
%
\bibitem{LAT} D. Toussaint, Nucl. Phys. Proc. Suppl. {\bf17} (1990) 
248;\\
I.M. Barbour, Nucl. Phys. Proc. Suppl. {\bf26} (1992) 22;\\
M.G. Alford, Nucl. Phys. Proc. Suppl. {\bf73} (1999) 161.
%
\bibitem{Hands} S.J. Hands, Nucl. Phys. Proc. Suppl. {\bf106} (2002) 459.
%
\bibitem{HW}
S.J. Hands and D.N. Walters, Phys. Lett. {\bf B548} (2002) 196.
%
\bibitem{RWA}
K.~Rajagopal and F.~Wilczek,
{\em in} {\em Handbook of QCD\/},
ch. 35, ed. M. Shifman (World Scientific, Singapore, 2001);\\
M.G.~Alford,
Ann.\ Rev.\ Nucl.\ Part.\ Sci.\  {\bf 51} (2001) 131.
%
\bibitem{BRHL} G.E. Brown and M. Rho, Phys. Rev. Lett. {\bf66} (1991) 2720;\\
T. Hatsuda and S.H. Lee, Phys. Rev. {\bf C46} (1992) 34.
%
\bibitem{HKK1} 
S.J. Hands, A. Koci\'c and J.B. Kogut, Ann. Phys. {\bf224} (1993)
29.
%
\bibitem{KRWP}
K.G. Klimenko, Z. Phys. {\bf C37} (1988) 457;\\
B. Rosenstein, B.J. Warr and S.H. Park, Phys. Rev. {\bf D39}
(1989) 3088.
%
\bibitem{HKK2} S.J. Hands, A. Koci\'c and J.B. Kogut, Nucl. Phys. {\bf B390} 
(1993) 355.
%
\bibitem{KS}
J.B. Kogut and C.G. Strouthos, Phys. Rev. {\bf D63}:054502 (2001).
%
\bibitem{HLM} S.J. Hands, B. Lucini and S.E. Morrison, Phys. Rev. {\bf
D65}:036004 (2002).
%
\bibitem{HM}
S.J. Hands and S.E. Morrison, Phys. Rev. {\bf D59}:116002
(1999).
%
\bibitem{BC} G. Baym and S.A. Chin, Nucl. Phys. {\bf A262} (1976) 527.
%
\bibitem{LeBellac} M. Le Bellac, {\em Thermal Field Theory\/}, (Cambridge
University Press, 1996).
%
\bibitem{Landau} 
L.D. Landau, 
Zh. Eksp. Teor. Fiz. {\bf30} (1956) 1058
(Sov. Phys. JETP {\bf3} (1956) 920).
%
\bibitem{Landau2}
E.M. Lifshitz and L.P. Pitaevskii, {\em Statistical Physics
(Part 2)\/} (Landau and Lifshitz Vol. 9) (Pergamon Press, Oxford 1980).
%
\bibitem{FW} A.L. Fetter and J.D. Walecka, 
{\em Quantum Theory of Many-Particle
Systems\/}, (McGraw-Hill, New York, 1971).
%
\bibitem{KT} J. Kapusta and T. Toimela, Phys. Rev. {\bf D37} (1988) 3731.
%
\bibitem{MEM}
C.R. Allton, J.E. Clowser, S.J. Hands, J.B. Kogut and C.G.
Strouthos,
Phys. Rev. {\bf D66}:094511 (2002).
%
\bibitem{LRR} K. Langfeld, H. Reinhardt and M. Rho, Nucl. Phys. {\bf A622}
(1997) 620; \\K. Langfeld, Nucl. Phys. {\bf A642} (1998) 96c.
%
\bibitem{SS} F. Sannino and W. Sch\"afer, Phys. Lett. {\bf B527} (2002) 142.
%
\bibitem{MNN} S. Muroya, A. Nakamura and C. Nonaka, Phys. Lett.
{\bf B551} (2003) 305.
%
\bibitem{Landau3} L.D. Landau,
Zh. Eksp. Teor. Fiz. {\bf32} (1957) 59
(Sov. Phys. JETP {\bf5} (1957) 101).
%
\bibitem{HKS} S.J. Hands, J.B. Kogut and C.G. Strouthos, Phys. Rev. {\bf
D65}:114507 (2002).
%
\bibitem{SoSt} D.T. Son and M.A. Stephanov, Phys. Rev. Lett. {\bf88} (2002)
202302.
%
\bibitem{AbSteg} M. Abramowitz and I.A. Stegun, {\em Handbook of Mathematical
Functions\/}, ch. 12 (Dover, New York, 1972).

\end{thebibliography}
\end{document}